\documentclass[twocolumn,prb,notitlepage,floats,superscriptaddress,amsmath,amssymb]{revtex4-1}

\usepackage[version=3]{mhchem} 
\usepackage{bm}
\usepackage[utf8]{inputenc}
\usepackage[T1]{fontenc}
\usepackage{graphicx}
\usepackage{upgreek}
\usepackage{color}
\usepackage{ulem}
\usepackage{hyperref} 
\hypersetup{colorlinks,citecolor=blue, filecolor=blue ,linkcolor=blue , urlcolor=blue, pdftex}
\usepackage{sidecap}
\usepackage{amssymb}

\def \FUW{Institute of Experimental Physics, Faculty of Physics, University of Warsaw, ul. Pasteura 5, 02-093 Warsaw, Poland}
\def \LNCMI{Laboratoire National des Champs Magn\'etiques Intenses, CNRS-UGA-UPS-INSA-EMFL, 25, avenue des Martyrs, 38042 Grenoble, France} 
\def \Taniguchi{International Center for Materials Nanoarchitectonics,
National Institute for Materials Science,  1-1 Namiki, Tsukuba 305-0044, Japan}
\def \Watanabe{Research Center for Functional Materials,
National Institute for Materials Science, 1-1 Namiki, Tsukuba 305-0044, Japan}
\def \Prague{Department of Condensed Matter Physics, Faculty of Mathematics and Physics, Charles University, Ke Karlovu 5, Praha 2 CZ-121 16, Czech Republic}

\begin{document}

\author{M.~Grzeszczyk}\email{magdalena.grzeszczyk@fuw.edu.pl}\affiliation{\FUW}
\author{J.~Szpakowski}\affiliation{\FUW}
\author{A. O. Slobodeniuk}\affiliation{\Prague}
\author{T.~Kazimierczuk}\affiliation{\FUW}
\author{M.~Bhatnagar}\affiliation{\FUW}
\author{T.~Taniguchi}\affiliation{\Taniguchi}
\author{K.~Watanabe}\affiliation{\Watanabe}
\author{P.~Kossacki}\affiliation{\FUW}
\author{M.~Potemski}\affiliation{\FUW}\affiliation{\LNCMI}
\author{A.~Babiński}\affiliation{\FUW}
\author{M.~R.~Molas}\email{maciej.molas@fuw.edu.pl}\affiliation{\FUW}

\title{The optical response of artificially twisted  \ce{MoS2} bilayers}

\date{\today}

\begin{abstract}
Two-dimensional layered materials offer the possibility to create artificial vertically stacked structures possessing an additional degree of freedom -- $the$ $interlayer$ $twist$. We present a comprehensive optical study of artificially stacked bilayers (BLs) \ce{MoS2} encapsulated in hexagonal BN with interlayer twist angle ranging from 0 to 60 degrees using Raman scattering and photoluminescence spectroscopies. It is found that the strength of the interlayer coupling in the studied BLs can be estimated using the energy dependence of indirect emission versus the A$_\textrm{1g}$-E$_\textrm{2g}^1$ energy separation. Due to the hybridization of electronic states in the valence band, the emission line related to the interlayer exciton is apparent in both the natural (2H) and artificial (62$^\circ$) \ce{MoS2} BLs, while it is absent in the structures with other twist angles. The interlayer coupling energy is estimated to be of about 50~meV. The effect of temperature on energies and intensities of the direct and indirect emission lines in \ce{MoS2} bilayers is also quantified.
\end{abstract}

\maketitle

\section{Introduction}
Two-dimensional (2D) van der Waals (vdW) crystals have emerged as a new generation of materials with extraordinary properties. For instance, widely studied semiconducting transition metal dichalcogenides (S-TMDs) transform from indirect- to direct-band gap, optically-bright semiconductors when thinned down to a monolayer (ML), which results in unique electronic structures and consequent optical properties~\cite{mak2010, AroraWSe2, AroraMoSe2, MolasWS2}. 
The family of 2D layered materials grows day by day, hugely expanding the scope of possible phenomena to be explored in two dimensions. Growing is also the number of possible vdW heterostructures that one can create. Such 2D materials currently cover a vast range of properties allowing potential applications in, $i.a.$ spintronic  devices~\cite{ciorciaro2020observation}, optoelectronics~\cite{sun2020lateral, shi2020ultrafast}, tunnel field-effect transistors~\cite{bandurin2017high, britnell2013resonant}, single-photon sources~\cite{koperski2015single, kern2016nanoscale}, and quantum information processing~\cite{kumar2015strain, branny2017deterministic}. 
Rapid advances in fabrication methods, like chemical vapor deposition (CVD) growth and mechanical exfoliation techniques, have contributed to increased interest in artificial stacking of different layered materials on top of each other. The simplistic approach of producing vertical vdW heterostructures without the constraints of crystal lattice mismatch enables integrating various 2D materials to create diverse systems with new electronic properties that are not present in pristine components. 
In addition to the selection of compounds in terms of their properties, a new degree of freedom has emerged: $the$ $twist$ $angle$ between stacked layers, which gives rise to the group of the so-called $twistronic$ $materials$~\cite{kar,feng}. The twist angle is responsible for the occurrence of moir\'e patterns, that leads to new and intriguing phenomena, like the formation of secondary Dirac points in graphene on hexagonal boron nitride (hBN)~\cite{ponomarenko2013cloning, hunt2013massive} or hybridized (moir\'e) excitons in vdW heterostructures formed by stacked two S-TMD MLs~\cite{rivera2015observation, zhang2018moire, Jin2019, Tran2019,Seyler2019, alexeev2019resonantly}. 

Raman scattering (RS) and photoluminescence (PL) spectroscopies are extensively used experimental techniques to characterize layered materials. Particularly, they can be used to determine the thickness of each \mbox{S-TMD} thin layer due to the energy dependence of phonon vibrations~\cite{lee2010anomalous, golasa2014resonant, yamamoto2014strong, Lui2015, grzeszczyk2016raman, kipczak2020, Holler2020, grzeszczyk2020breathing} as well as to the direct-indirect band-gap transformation~\cite{mak2010, AroraWSe2, AroraMoSe2, MolasWS2}. 

In this work, we investigate interlayer interactions in high-quality artificially stacked twisted \ce{MoS2} bilayers (BLs) encapsulated in hBN flakes using the RS and PL spectroscopies. Our results indicate that the interlayer coupling can be determined by the comparison of the emission energies of the indirect transition versus the energy separation between two basic intralayer phonon modes (A$_\textrm{1g}$ and E$_\textrm{2g}^1$). The origin of the apparent emission due to the interlayer excitons in both the natural (2H) and artificial (62$^\circ$) \ce{MoS2} BLs and its absence for the BLs with other twist angles is associated with hybridization of electronic states in the valence band. 
We also investigate the evolution of energies and intensities of the  emission lines due to direct and indirect transitions with temperature ranging from 5 K to 300 K.

\section{Results}
\subsection{Atomic structures of twisted \ce{MoS2} bilayers}
\begin{figure}[t]
    \centering
    \includegraphics[width=\linewidth]{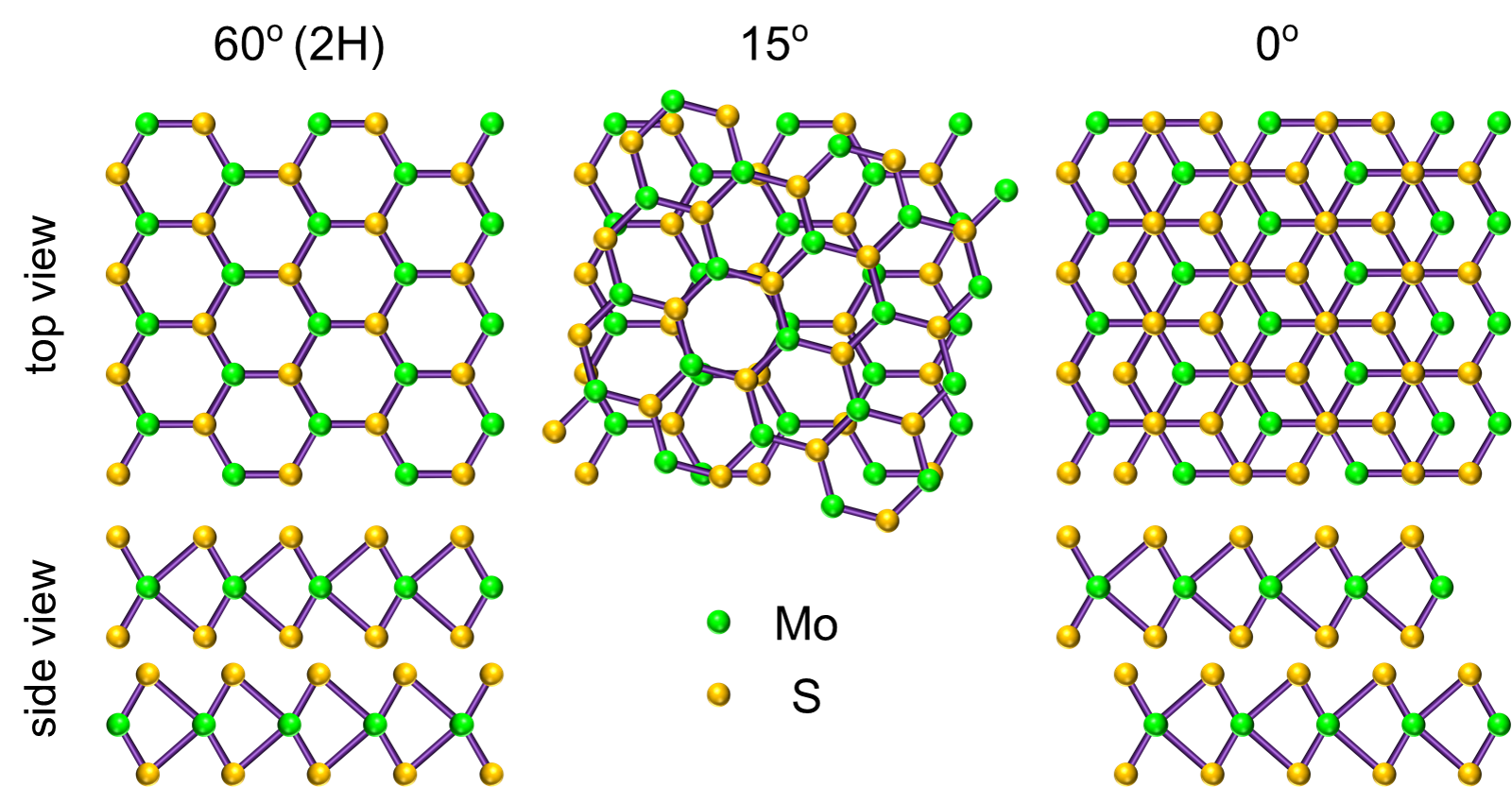}
    \caption{Atomic structure and stacking order of  \ce{MoS2} bilayers for different twist angles: 60$^\circ$, 15$^\circ$ and 0$^\circ$. Note that the side view demonstrates the energetically favourable structures for angles of 60$^\circ$ and 0$^\circ$.}
    \label{fig:stacking}
\end{figure}

\ce{MoS2} belongs to the family of S-TMDs with chemical formula \ce{MX2} where M=Mo or W and X=S, Se or Te, which the most common crystallographic structure is a hexagonal phase. In that case, X--Mo--X atoms in a monolayer (1~L) of \ce{MoX2} are arranged in a trigonal prismatic structure, which does not exhibit inversion symmetry. A BL formed by the stacking of two MLs exhibits an additional degree of freedom -- the twist angle between the layers. Due to the hexagonal symmetry of a \ce{MX2} ML, the different arrangements are characterized by twist angles ranging from 0$^\circ$ to 60$^\circ$. The schematic illustration of three patterns of \ce{MoS2} BLs with the twist angles of 60$^\circ$, 15$^\circ$, and 0$^\circ$ are presented in Fig.~\ref{fig:stacking}. The most stable structures, which exhibit the strongest coupling, are found for twist angles equal to 0$^\circ$ and 60$^\circ$ ascribed correspondingly to the 2H and 3R stackings \cite{liu2014evolution, Baren2019}. We do not use the label 3R for the 0$^\circ$ BL (see Fig.~\ref{fig:stacking}), because the 3R unit cell involves the atoms from three consecutive layers (2H unit cell is composed of two layers). While both the 2H and 3R polytypes are accessible in natural and CVD-grown \ce{MoS2} multilayers~\cite{Baren2019}, the BLs with other twist angles can only be constructed by the artificial stacking of individual monolayers. It is important to point out that the inversion symmetry is restored in the 2H BL, while the zero-twist angle BL has neither inversion nor in-plane mirror.

\subsection{Raman scattering spectroscopy}
\begin{figure}[t!]
    \centering
    \includegraphics[width=\linewidth]{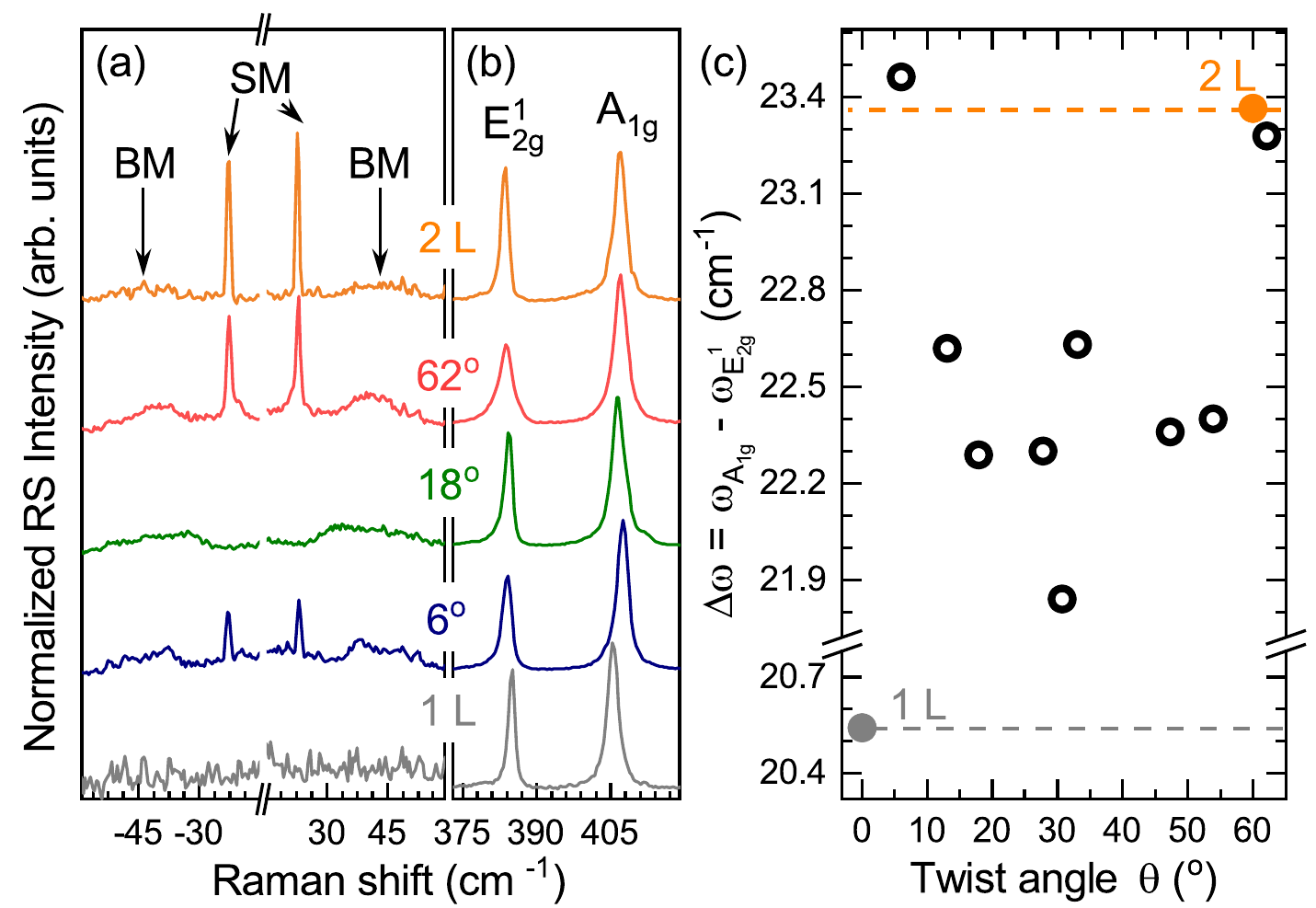}
    \caption{Normalized room-temperature Raman scattering spectra of thin layers MoS$_2$: monolayer (1~L), bilayer (2~L) and selected homobilayers with twist angles of 6$^\circ$, 18$^\circ$ and 62$^\circ$ in the energy range of (a) interlayer and (b) intralayer vibrations measured under excitation of 2.41~eV laser light. The spectra are normalized to the intensity of the A$_\textrm{1g}$ peak. (c) The energy separation between the A$_\textrm{1g}$ and E$_\textrm{2g}^1$ peaks, $i.e.$ $\Delta  \omega=\omega_{\textrm{A}_\textrm{1g}}-\omega_{\textrm{E}_\textrm{2g}^1}$, as a function of twist angle ($\theta$) for all the studied twisted homobilayers. The corresponding $\Delta \omega$ for 1~L and 2~Ls, represented by solid symbols, are also shown. }
    \label{fig:raman}
\end{figure}

In order to investigate the interlayer coupling in the twisted BLs and therefore to distinguish their pristine properties, we compare their RS and PL spectra with those obtained for 1~L and 2~L exfoliated from 2H bulk \ce{MoS2}. Note that the twist angle between MLs forming BLs was determined using the second-harmonic generation (SHG) technique, which permits to reveal the crystallographic orientation of individual MLs \cite{Hsu2014}. The measured RS spectra of the selected BLs with twist angles of 62$^\circ$, 18$^\circ$, and 6$^\circ$ accompanied with the ones of 1~L and 2~L \ce{MoS2} are presented in Fig.~\ref{fig:raman}(a) and (b). 

First, we focus on the low-energy range of the RS spectra, presented in Fig.~\ref{fig:raman}. There are no modes related to rigid vibrations in 1~L affirming their interlayer nature. For 2~L, two modes are observed at $\sim$23~cm$^{-1}$ and $\sim$44~cm$^{-1}$, which are associated with the in-plane shear (SM) and out-of-plane breathing (BM) vibrations of rigid layers, respectively~\cite{zhang2013, zhao2013, Baren2019}. In twisted BLs with angles of 62$^\circ$ and 6$^\circ$, the SM and BM modes are also apparent. This indicates that the coupling between adjacent layers in these two BLs is relatively strong due to their resemblance to the natural 2H and 3R stackings. Note that the relatively smaller intensity of the SM peak in the 6$^\circ$ BLs as compared with the 2H and 62$^\circ$ ones can be attributed to the smaller interlayer bond polarizability for the 3R polytype~\cite{puretzky2015low}. This effect was applied to identify the stacking order of S-TMD layers~\cite{puretzky2015low, Holler2020}. On the other hand, only the BM-related peak can be observed for the sample with the 18$^\circ$ angle as well as for other BLs with angles in the range from $\sim$10$^\circ$ to $\sim$56$^\circ$ (data not shown). 
The higher energy range of all RS spectra shown in Fig.~\ref{fig:raman}(b) is dominated by two phonon modes ascribed to the in-plane ($\mathrm{E^1_{2g}}$) and out-of-plane ($\mathrm{A_{1g}}$) intralayer vibrations.  In few layers 2H \ce{MoS2}, the $\mathrm{E^1_{2g}}$ ($\mathrm{A_{1g}}$) phonon mode experiences a red (blue) shift with increasing the layer thickness. The energy separation between those modes, $i.e.$ $\Delta \omega=\omega_{\textrm{A}_\textrm{1g}}-\omega_{\textrm{E}_\textrm{2g}^1}$, is a useful tool to determine the number of \ce{MoS2} layers~\cite{lee2010anomalous}. As can be seen in the Figure, the $\Delta  \omega$ energy difference varies for different twist angles ($\Theta$), which is summarized for all samples in Fig.~\ref{fig:raman}(c). The $\Delta \omega$ values are largest for samples with a twist angle of 6$^\circ$ and 62$^\circ$ which correspond well to natural 2H and 3R BLs~\cite{Lee2016}. In the intermediate cases (6$^\circ$<$\Theta$<60$^\circ$), the $\Delta \omega$ falls lower than for 2~L \ce{MoS2}, but it is considerably larger than in 1~L. The $\Delta \omega$ can be also used to characterize the effective interlayer mechanical coupling strength or, in other words, the distance between layers in twisted BLs~\cite{liu2014evolution}. In such case, we can conclude that the structures with the twist angle equal to 6$^\circ$ and 62$^\circ$ are characterized by the strongest coupling (the smallest interlayer distance), while the coupling strength is weaker (the interlayer distance is bigger) for other BLs with the twist angles substantially different from 0 and 60 degrees. 

Note that our RS results are consistent with previous works devoted to the twisted \ce{MoS2} BLs exfoliated on Si/SiO$_2$ substrates~\cite{liu2014evolution, zande2014, huang2016, Lin2018, Liao2020}.

\subsection{Photoluminescence spectroscopy}

\begin{figure}
    \centering
    \includegraphics[width=\linewidth]{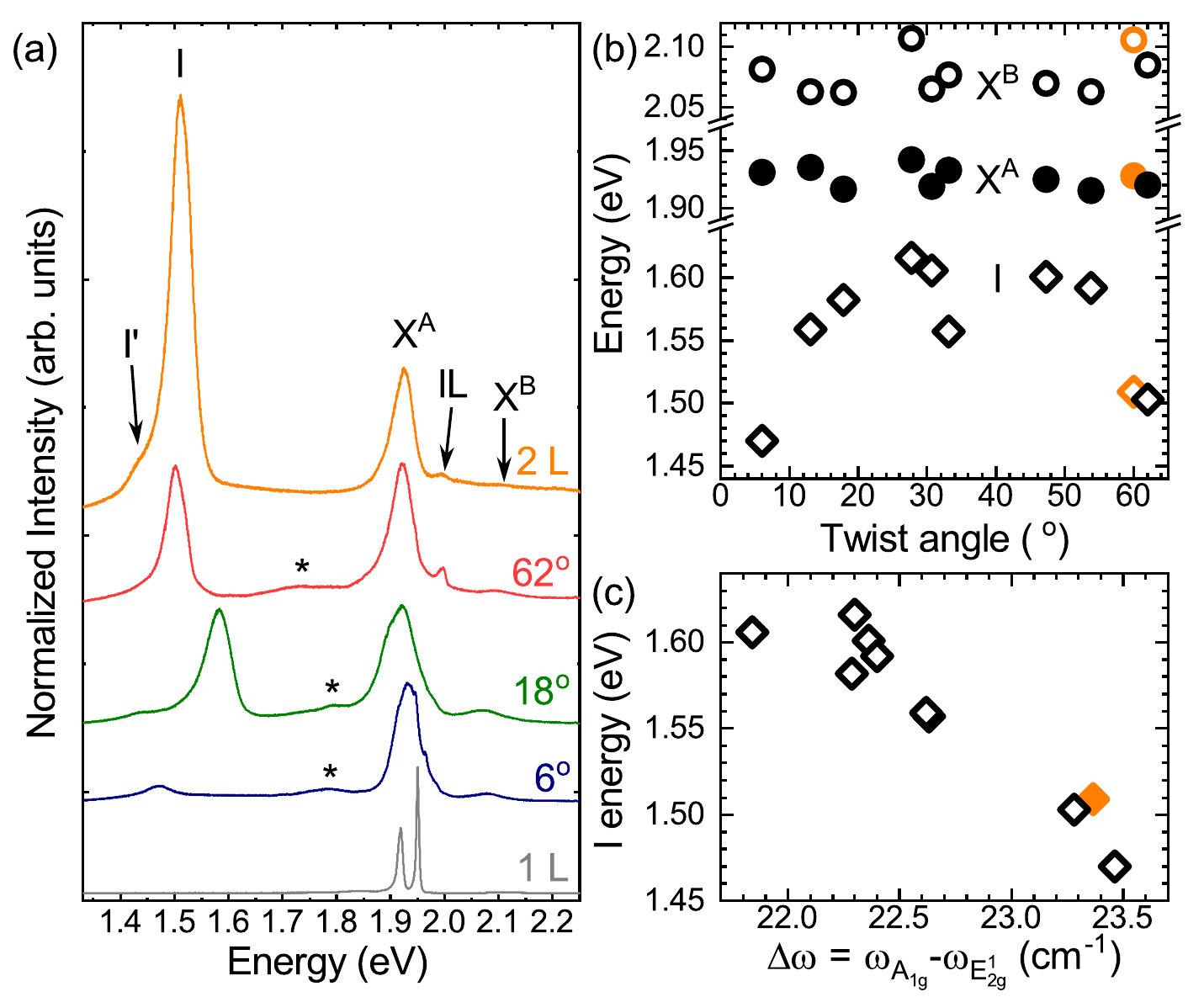}
	    \caption{(a) Normalized low-temperature ($T$=5~K) photoluminescence spectra of thin MoS$_2$ layers: monolayer (1~L), bilayer (2~L) and selected homobilayers with twist angles of 6$^\circ$, 18$^\circ$ and 62$^\circ$ measured under excitation of 2.41~eV laser light. The spectra are normalized to the intensity of the X$^\textrm{A}$ emission. (b) The emission energy of the direct (X$^\textrm{A}$ nad X$^\textrm{B}$) and indirect (I) transitions as a function of twist angle ($\theta$) for all the studied twisted homobilayers. (c) The evolution of the indirect emission energy, $i.e.$ I, versus the energy separation between the A$_\textrm{1g}$ and E$_\textrm{2g}^1$ peaks, $i.e.$ $\Delta \omega=\omega_{\textrm{A}_\textrm{1g}}-\omega_{\textrm{E}_\textrm{2g}^1}$. The corresponding emission energies for the 2~L are also indicated with solid orange points.}
    \label{fig:PL_low}
\end{figure}

Previous investigation of PL spectra of artificially twisted BLs \ce{MoS2} were carried out at room temperature~\cite{zande2014, liu2014evolution, Liao2020}. The PL spectroscopy at low temperature may provide however a more accurate analysis, which is due to the much lower emission linewidths~\cite{MolasWS2,Du2018}. The PL spectra of the selected BLs with twist angles of 62$^\circ$, 18$^\circ$, and 6$^\circ$ accompanied with those of 1~L and 2~L \ce{MoS2} measured at low temperature ($T$=5~K) are shown in Fig.~\ref{fig:PL_low}(a). Similarly to the aforementioned analysis of the RS, we begin with an examination of the PL spectra of the natural 1~L and 2~L \ce{MoS2}. The 1~L spectrum consists of two narrow emission lines apparent in the vicinity of the optical band gap (so-called A exciton), which can be ascribed to the neutral and charged excitons in accordance with previous reports~\cite{Cadiz2016, Cadiz2017, MolasAcid}. In contrast, the PL spectrum of the 2~L \ce{MoS2} is composed of two distinct emission bands: (i) the transitions in the vicinity of the direct A (X$^\textrm{A}$) and B (X$^\textrm{B}$) excitons formed at the K$^\pm$ points of the Brillouin zone (BZ), which are observed correspondingly at about 1.93~eV and 2.07~eV; (ii) the significantly much intense transitions, denoted as I and I', apparent at about 1.5~eV, which are ascribed correspondingly to an indirect recombination process between the $\Lambda$ and K point in the conduction band (CB) and $\Gamma$ points in the valence band (VB) of the BZ~\cite{Tongay2012, liu2014evolution, Du2018}. The PL spectra of the artificial twisted BLs also comprise two direct- and indirect-related bands. While the energies of direct transitions are hardly affected by the twist angles, the energies of the indirect ones change significantly by about 150~meV. The low-intensity emission bands, denoted with * and apparent between the aforementioned I and X$^\textrm{A}$ lines, can be described as a remaining part of the defect-related emission, which is reported commonly for 1-L \ce{MoS2} exfoliated on Si/SiO$_2$ substrates~\cite{Cadiz2016, Cadiz2017, MolasAcid}.

To appreciate the effect of the twist angle on the PL spectra, the energy evolution of the X$^\textrm{A}$, X$^\textrm{B}$ and I transitions for all the studied samples as a function of twist angle is presented in Fig.~\ref{fig:PL_low}(b). As can be seen in the Figure, there is no effect of the twist angle on the X$^\textrm{A}$ and X$^\textrm{B}$ energies. This reflects a pure two-dimensional character of both the A and B excitons, $i.e.$ these complexes are distributed spatially within a single layer even in a bulk form of \ce{MoS2}~\cite{Koperski2017, AroraBULK2017, AroraBULK2018}. For the indirect transitions, we focus only on the I emission lines as they are observed for all studied samples. The emission reaches the lowest energies for border cases (6$^\circ$ and 62$^\circ$), which are similar to the value obtained for the natural 2~L \ce{MoS2}. For BLs with other twist angles, the I energies are substantially higher by about 150~meV and they are almost independent of twist angle. This effect can be associated with changes in the interlayer distance between MLs forming the studied BLs as a function of twist angle, as it was reported previously in Refs.~\citenum{liu2014evolution, zande2014}. 

As the energy separation between the A$_\textrm{1g}$ and E$_\textrm{2g}^1$ peaks ($\Delta \omega$) can be considered as a probe of the interlayer distance in the studied artificial \ce{MoS2} BLs, one can plot the energy dependence of the indirect transition I as a function of $\Delta \omega$ in Fig.~\ref{fig:PL_low}(c). The presented results can be divided into two groups: (i) for structures with twist angle close to 0$^\circ$ and 60$^\circ$, $i.e.$ $\Delta\omega \sim$23~cm$^{-1}$, the I emission energy equals approx. 1.5~eV; (ii) for samples with twist angle significantly different from 0$^\circ$ and 60$^\circ$, $i.e.$ $\Delta \omega\sim$22~cm$^{-1}$, the energy dispersion of the I line is of about 80~meV centered at $\sim$1.58~eV. These results are consistent with previously obtained for stacked triangle-shaped \ce{MoS2} BLs grown on Si/SiO$_2$ substrate using CVD technique\cite{liu2014evolution}. Summarizing, we can assume that the analysis of the indirect emission versus the $\Delta \omega$ can be used as a useful tool to probe the interlayer distance in the twisted homo-BLs. This allows to distinguish between two cases of twist angles, which are in the vicinity of 0$^\circ$ and 60$^\circ$ or the twist angle is in-between.

\begin{figure}[t]
    \centering
    \includegraphics[width=\linewidth]{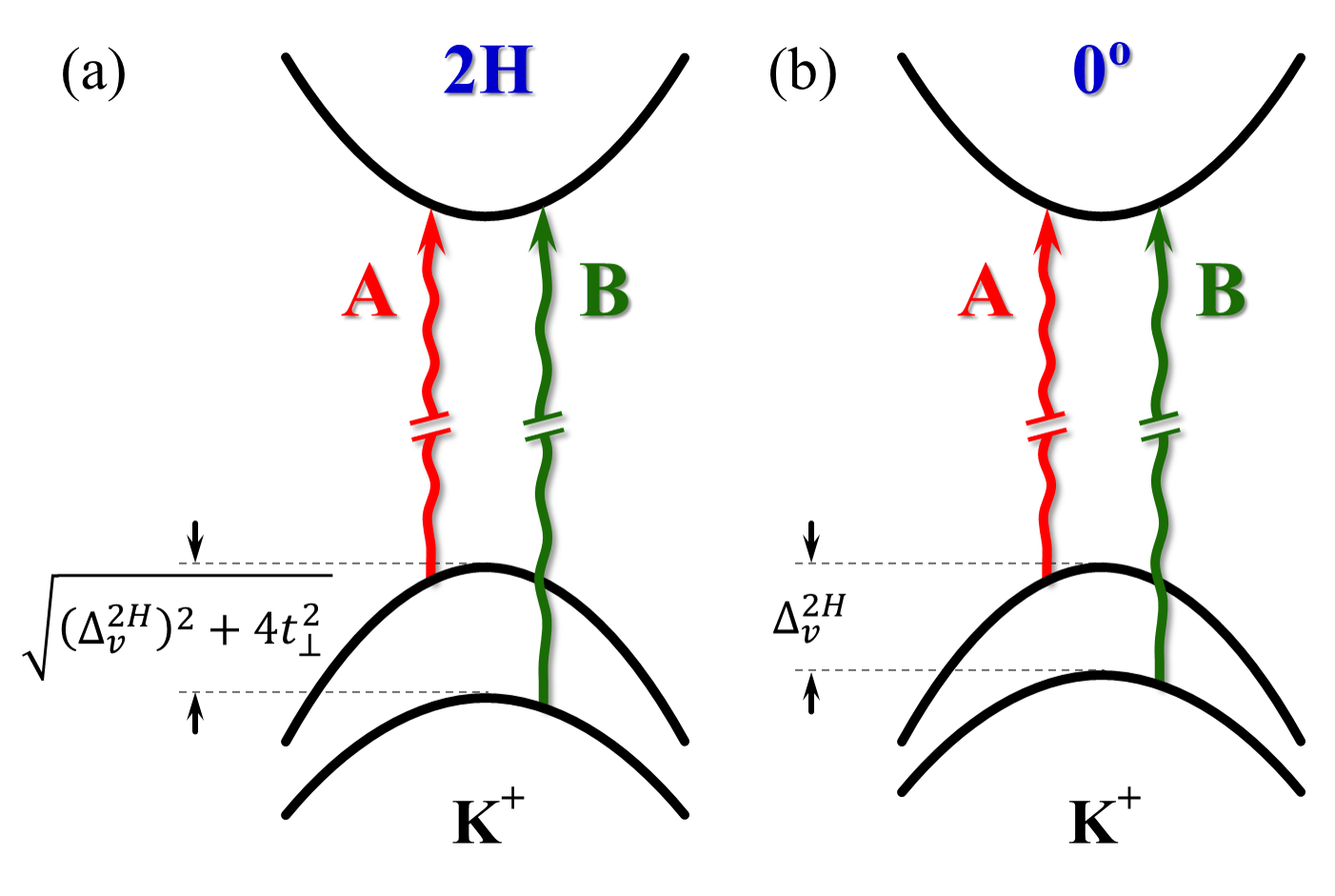}
    \caption{Diagram of relevant subbands in the CB and VB at the K$^+$ point of the Brillouin zone in the bilayers \ce{MoS2} with twist angle of (a) 60$^\circ$ (2H) and (b) 0$^\circ$. The red and green wavy lines show the A and B exciton transitions. The $t_\perp$ and $\Delta^\textrm{2H}_v$ represent the interlayer hopping term and the spin-orbit VB splitting without interlayer coupling, respectively.}
    \label{fig:scheme}
\end{figure}

While the X$^\textrm{A}$ and X$^\textrm{B}$ lines are observed for all studied samples, there is an additional emission line, labelled IL, apparent between the former ones for the 2~L and 62$^\circ$ samples, which emission was not reported so far. 
According to recent results of reflectance contrast (RC) experiments carried out on natural 2H \ce{MoS2} BLs at low temperature~\cite{Slobodeniuk2019, Gerber2019, Paradisanos2020}, that line can be associated with the recombination of the so-called interlayer A exciton.  The IL line originates from the hybridization of electronic states in the VB of the 2H-stacked BL due to the interlayer coupling (see Ref.~\citenum{Slobodeniuk2019} for details). The strength of the related VB coupling is described by the interlayer hopping term, $t_\perp$. Based on a \textbf{kp} model of 2H BLs in the vicinity of $\mathrm{K}^\pm$ points~\cite{Slobodeniuk2019}, the $t_\perp$ parameter is given by:
\begin{equation}
t_\perp=\frac{1}{2}\sqrt{(\Delta^\textrm{2H}_\textrm{A-B})^2-(\Delta^\textrm{2H}_v)^2},
\end{equation}
where $\Delta^\textrm{2H}_\textrm{A-B}$ is the A-B energy difference for natural 2H BL and $\Delta^\textrm{2H}_v$ represents the spin-orbit VB splitting without interlayer coupling. Note that the detailed investigation of the band structure of the 2H and 0$^\circ$ BLs in the vicinity of the K$^\pm$ point of the BZ is performed in Appendices~\ref{app:0_stacking} and \ref{app:2H_stacking}. According to that analysis, we assume that: (i) the $\Delta^\textrm{2H}_v$ is an order of magnitude larger than its counterpart in the CB $\Delta^\textrm{2H}_c$~\cite{Kormanyos2015}; (ii) the binding energies of the A and B excitons are comparable; (iii) the $\Delta^\textrm{2H}_v$ can be roughly approximated by the A-B energy difference for the BL with twist angle of 0$^\circ$. Consequently, the diagram of relevant subbands at the K$^+$ point of the BZ in the \ce{MoS2} BLs with twist angle of 60$^\circ$ (2H) and 0$^\circ$ is shown in Fig.~\ref{fig:scheme}. As can be appreciated from the Figure, the $t_\perp$ parameter can be evaluated using the energy separation between the A and B excitons measured in the 60$^\circ$ (2H) and 0$^\circ$ BLs. The extracted $\Delta^\textrm{2H}_\textrm{A-B}\sim$180~meV (the corresponding $\Delta^{62^\circ}_\textrm{A-B}$ is about 10 meV smaller, which suggests the lower strength of interlayer coupling in twisted BL) and $\Delta^{6^\circ}_\textrm{A-B}\sim$150~meV are in very good agreement with the reported ones for natural BLs~\cite{Slobodeniuk2019, Gerber2019, Paradisanos2020}. Using these values, the extracted interlayer hopping parameter ($t_\perp$) is found to be on the order of 50~meV, which agrees very well with the value obtained using the RC experiment performed on the \ce{MoS2} BLs grown using CVD technique and encapsulated in hBN flakes (49~meV)~\cite{Paradisanos2020}. 

\begin{figure}[t]
    \centering
    \includegraphics[width=1\linewidth]{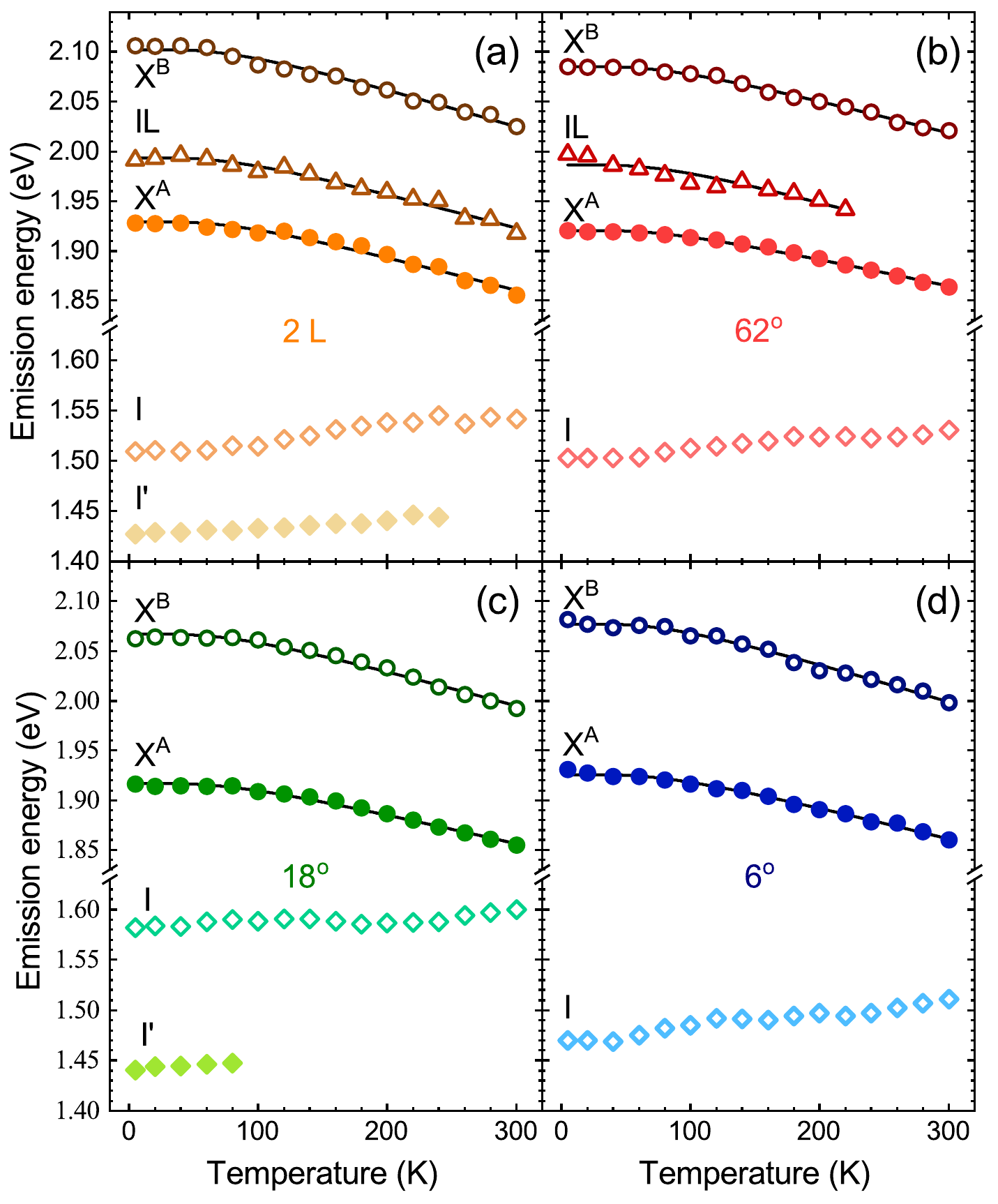}
    \caption{Temperature evolution of the energies of the transitions extracted from PL spectra measured on MoS$_2$: (a) bilayer (2~L) and selected homobilayers with twist angles of (b) 62$^\circ$, (c) 18$^\circ$ and (d) 6$^\circ$. The solid black curves are fits to the data obtained with the aid of the O’Donnell formula.}
    \label{fig:PL_temp}
\end{figure}

The energy evolution of both the direct (X$^\textrm{A}$, X$^\textrm{B}$, and IL) and indirect (I and I') lines as a function of temperature measured on BL (2~L) and selected homobilayers with twist angles of 6$^\circ$, 18$^\circ$ and 62$^\circ$ is shown in Fig.~\ref{fig:PL_temp}. As can be appreciated in the Figure, all lines within the direct- and indirect-related transitions located at $\sim$2.0~eV and $\sim$1.5~eV are characterized by the same type of evolution upon increasing temperature, $i.e.$ redshift and blueshift, respectively. The X$^\textrm{A}$, X$^\textrm{B}$, and IL peaks redshift when temperature is increased from 5~K to 300~K. This can be associated with the reduction of the direct band gap resulting from the temperature expansion of those layers in lateral directions. Consequently, these evolutions can described by the relation proposed by O’Donnell $et$ $al.$~\cite{Odonnell}, which expresses the temperature dependence of the band gap in terms of the average energy of acoustic phonons involved in the electron-phonon interaction \(\langle \hbar \omega \rangle\). The relation reads $E(T)=E_0-S \langle \hbar \omega \rangle [ \textrm{coth} ( \langle \hbar \omega \rangle / 2k_{\textrm{B}} T )-1 ]$,  where $E_0$ stands for the band gap at absolute zero temperature, $S$ is the coupling constant, and $k_\textrm{B}$ denotes the Boltzmann constant. We found that $\langle \hbar\omega \rangle$ stayed on nearly the same level, $\sim21$ meV, for the X$^\textrm{A}$, X$^\textrm{B}$, and IL transitions in 2H BL~\cite{Tongay2012}. As a consequence, we kept it fixed during fittings of all the experimental data shown in Fig.~\ref{fig:PL_temp}. It can be seen that the fitted curves correctly reproduce the energies of excitonic lines (see Fig.~\ref{fig:PL_temp}), which suggests that binding energies of investigated complexes do not depend on temperature. Interestingly, the overall redshift of X$^\textrm{A}$ lines are of about 60--70~meV, while the corresponding shift for the \ce{MoS2} BL exfoliated on Si/SiO$_2$ substrate was found to be of almost 90~meV~\cite{Du2018}. Simultaneously, the crystal expansion across the layers leads to a larger separation between the layers, which results in the blueshift of the indirect band gap. The I emission lines experience almost monotonic blueshifts in the range of about 20--40~meV, while the shift of $\sim$80~meV is reported in Ref.~\citenum{Du2018}. Both these results indicate that the hBN encapsulation of the BLs induces much smaller expansion of the flake in both directions (in lateral directions and across the layers) with increasing temperature as compared with the BLs exfoliated on Si/SiO$_2$ substrate~\cite{Du2018}.

\begin{figure}[t]
    \centering
    \includegraphics[width=1\linewidth]{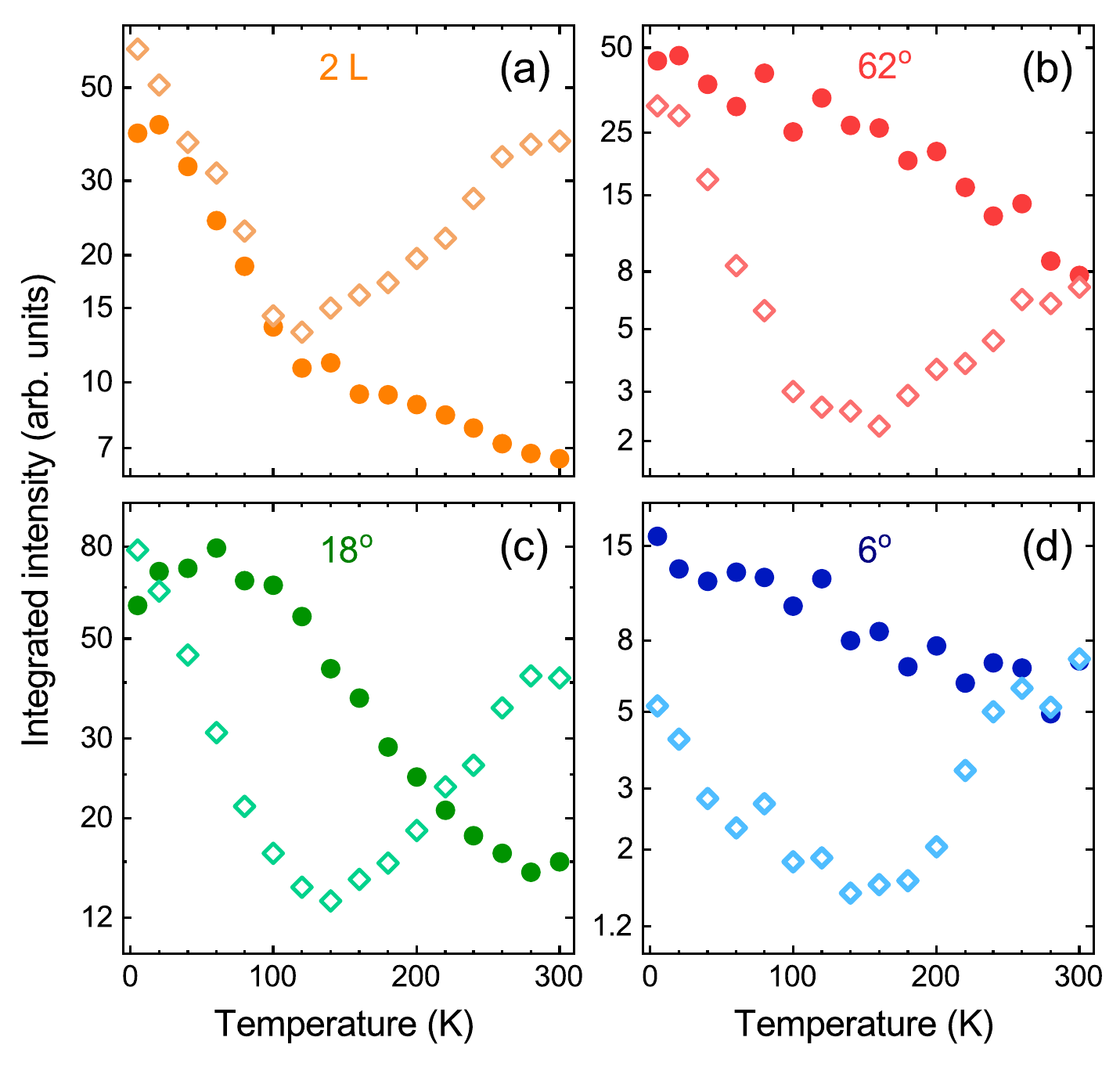}
    \caption{Temperature evolution of the integrated intensities of the (solid circular points) direct- and (open diamond points) indirect-related emissions extracted from PL spectra measured on \ce{MoS2}: (a) bilayer (2~L) and selected homobilayers with twist angles of (b) 62$^\circ$, (c) 18$^\circ$ and (d) 6$^\circ$. Note that the vertical scales are logarithmic.}
    \label{fig:PL_int_temp}
\end{figure}

The temperature evolution of the integrated intensities of the direct- and indirect-related transitions is presented in Fig.~\ref{fig:PL_int_temp}. It can be seen that the total intensities of all direct transitions mostly decrease monotonically with increasing temperature. The observed trends are in agreement with previous reports for direct emissions apparent in \ce{MoS2} ML~\cite{Tongay2012,Wang2015} and probably result from the competition between the efficiencies of the radiative and non-radiative recombination channels. Surprisingly, the temperature dependencies of the corresponding indirect transition intensities are clearly non-monotonic. First, quick reductions of the related intensities is observed from 5~K to about 100~K-150~K, which can be associated with increase of kinetic energies of excitons and interplay between radiative and non-radiative processes. Then it is followed by slower increase of the intensities up to room temperature, which can be explained in terms of the increased population of phonons at higher temperatures. 

\section{Summary}
A systematic investigation of optical properties of hBN-encapsulated artificially stacked \ce{MoS2} BLs with varying interlayer twist angle has been conducted. It has been shown that RS spectroscopy is a sensible tool for determining the interlayer coupling and spacing in the BL. The strongest coupling has been observed in \ce{MoS2} homobilayers with twist angles of 6$^{\circ}$ and 62$^{\circ}$, which reproduce well the structure of 2H and 3R polytypes found in natural and CVD-grown \ce{MoS2} multilayers. Increasing the twist angle of homobilayers from 6$^{\circ}$ to $\sim30^{\circ}$ leads to an increase in the interlayer spacing, and thus a decrease in the interlayer coupling. This effect can be observed as the absence of the low-energy interlayer phonon modes in the RS spectrum as well as the energy shift of $\sim140$~meV for the indirect transition between K-$\Gamma$ points of the Brillouin zone. The PL of artificially assembled homobilayers, in comparison with the natural 2~L \ce{MoS2}, features a single line associated with indirect emission, but its energy varies considerably depending on the interlayer twist. We conclude that gaining a new degree of freedom by inter-layer twisting of artificially assembled flakes permits the control over the energy of the indirect transition, This offers further insight in few-layer 2D systems and can be a useful tool for future device applications.

\section*{Methods}
The investigated \ce{MoS2} BLs, MLs, and hBN flakes were fabricated by two-stage PDMS-based mechanical exfoliation of bulk crystals. An unoxidized Si wafer was used as a substrate. In order to ensure the best quality of the substrate surface, they were annealed at 200$^{\circ}$C and kept on hot-plate until the first non-deterministic transfer of h-BN flakes. Subsequent layers were transferred deterministically, to reduce inhomogeneities between each transfer sample were annealed. 
The complete structures were annealed at 160$^{\circ}$C for 1.5 hours to ensure the best layer-to-layer and layer-to-substrate adhesion and to eliminate a substantial portion of air pockets on the interfaces between the constituent layers.

The SHG measurements performed to identify the relative twist angle of the exfoliated S-TMD MLs forming homobilayers were taken at $T=300$~K using a home-built setup with a femtosecond Ti:Sapphire laser with excitation at 800~nm (1.55~eV). For each measurement, the laser light had a typical incident power of 500~$\upmu$W, was linearly polarized, and focused to a spot size of 1~$\upmu$m by a 50x objective lens. A set of a motorized half-wave plate and a fixed linear polarizer were used to analyse an SHG signal, which was detected by a Si avalanche photodiode.

The PL and RS measurements were performed using \mbox{$\lambda$=515~nm} (2.41~eV) radiation from continuous-wave Ar-ion or diode lasers. For the PL and RS experiments at room temperature ($T=300$~K), the excitation light was focused through a 100x long-working distance objective with a 0.55 numerical aperture (NA) producing a spot of about 1~$\upmu$m diameter. 
The signal was collected via the same microscope objective, sent through a 1~m monochromator, and then detected by using a liquid nitrogen-cooled charge-coupled device (CCD) camera. To detect low-energy RS up to about $\pm$10~cm$^{-1}$ from the laser line, a set of Bragg filters was implemented in both excitation and detection paths. The temperature-dependent PL measurements were performed using an analogous setup with small modifications (a 50x long-working distance objective and a 0.5~m monochromator). Moreover, the studied samples were placed on a cold finger in a continuous flow cryostat mounted on $x$-$y$ motorized positioners. The excitation power focused on the sample was kept at 200~$\upmu$W during all measurements to avoid local heating.

\begin{acknowledgments}
The work has been supported by the National Science Centre, Poland (grant no. 2017/27/B/ST3/00205, 2017/27/N/ST3/01612, 2018/31/B/ST3/02111), EU Graphene Flagship project (no.785219), the ATOMOPTO project (TEAM programme of the Foundation for Polish Science, co-financed by the EU within the ERD-Fund), and the CNRS via IRP "2DM" project. The Polish participation in EMFL is supported by the DIR/WK/2018/07 grant from Polish Ministry of Science and Higher Education. K. W. and T. T. acknowledge support from the Elemental Strategy Initiative conducted by the MEXT, Japan, (grant no. JPMXP0112101001), JSPS KAKENHI (grant no. JP20H00354), and the CREST (JPMJCR15F3), JST.
\end{acknowledgments}

\appendix
\section{Bilayer with $60^\circ$-angle alignment}
\label{app:2H_stacking}

We model a S-TMD bilayer with $60^\circ$-angle alignment as a pile of two monolayers (top and bottom), placed in parallel to $xy$ plane. We define the positions of metal and chalcogen atoms of the bottom layer as
\begin{align}
\mathbf{R}_{mn}^{M,b}&=\mathbf{a}_1m+\mathbf{a}_2n+\mathbf{t}_M=\mathbf{R}_{mn}+\mathbf{t}_M, \\
\mathbf{R}_{mn}^{X_\pm,b}&=\mathbf{a}_1m+\mathbf{a}_2n+\mathbf{t}_X\pm\eta\mathbf{e}_z=
\mathbf{R}_{mn}+\mathbf{t}_X\pm \eta\mathbf{e}_z,
\end{align}
respectively. Here we introduced the in-plane primitive lattice vectors of length $a_0$
\begin{equation}
\mathbf{a}_1=\frac{a_0}{2}(\mathbf{e}_x+\sqrt{3}\mathbf{e}_y), \quad
\mathbf{a}_2=\frac{a_0}{2}(\mathbf{e}_x-\sqrt{3}\mathbf{e}_y),
\end{equation}
the pair of integer numbers $(m,n)$, the short notation for the $(m,n)$-th lattice vector $\mathbf{R}_{mn}=\mathbf{a}_1m+\mathbf{a}_2n$ and unit vectors of Cartesian coordinate system
$\mathbf{e}_x,\mathbf{e}_y,\mathbf{e}_z$. Vectors
\begin{equation}
\mathbf{t}_M=\frac{a_0}{2}\Big(\mathbf{e}_x+\frac{1}{\sqrt{3}}\mathbf{e}_y\Big), \quad
\mathbf{t}_X=\frac{a_0}{2}\Big(\mathbf{e}_x-\frac{1}{\sqrt{3}}\mathbf{e}_y\Big)
\end{equation}
define in-plane positions of the metal and chalcogen atoms within a unit cell of S-TMD monolayer, respectively.
Vectors $\pm\eta\mathbf{e}_z$ with $\eta>0$ define the out-of-plane positions of chalcogen atoms in the unit cell.
We also introduce the primitive vectors of reciprocal lattice
\begin{equation}
\mathbf{b}_1=\frac{2\pi}{a_0}\Big(\mathbf{e}_x+\frac{1}{\sqrt{3}}\mathbf{e}_y\Big), \quad
\mathbf{b}_2=\frac{2\pi}{a_0}\Big(\mathbf{e}_x-\frac{1}{\sqrt{3}}\mathbf{e}_y\Big).
\end{equation}
They satisfy the orthogonality property $\mathbf{a}_j\mathbf{b}_k=2\pi\delta_{jk}$, where $\delta_{jk}$ is the Cronecker delta.

The top lattice of the bilayer can be obtained from the bottom one as a result of the shift along $z$ direction on some distance $l$ with subsequent rotation around $Oz$ axis on $180^\circ$ degree. Then, the positions of the metal and chalcogen atoms of the top lattice become 
\begin{align}
\mathbf{R}_{mn}^{M,t}&=\mathbf{R}_{mn}+l\mathbf{e}_z+\mathbf{t}_X, \\
\mathbf{R}_{mn}^{X_\pm,t}&=\mathbf{R}_{mn}+l\mathbf{e}_z+\mathbf{t}_M\pm \eta\mathbf{e}_z.
\end{align}
Note that all of the chalcogen atoms of bottom layer are aligned with the metal atoms of the top layer (and vice versa) along $z$-direction. Hence, the unit cell of the bilayer contains twice more atoms than in 1~L. The positions of metal and chalcogen atoms within the unit cell are defined by vectors
$\{\mathbf{t}_M, \mathbf{t}_X+l\mathbf{e}_z\}$ and
$\{\mathbf{t}_X\pm \eta\mathbf{e}_z, \mathbf{t}_M\pm\eta\mathbf{e}_z +l\mathbf{e}_z\}$, respectively.
This arrangement of atoms is called 2H-stacking and corresponds to thermodynamically stable form of all S-TMD crystals with any number of layers including bulk.

The bilayer possesses $C_3$ rotation symmetry (with $Oz$ line as the rotational axis) and mirror symmetry $P:x\leftrightarrow -x$ (the mirror's plane is $yz$-plane). Therefore, the crystal has the same hexagonal Brillouin zone as the Brillouin zone of the bottom layer. Hence, it is convenient to use the known Bloch states of the monolayer as the basis states. Namely we are interested in conduction (\textit{c}) and valence (\textit{v}) band states at the $\mathrm{K}^\pm$ points. To this end, we introduce the vectors $\pm\mathbf{K}=\pm(\mathbf{b}_1+\mathbf{b}_2)/3$, which define the position of the $\mathrm{K}^\pm$ points in the reciprocal space, respectively. In further we will use the notation $\pm\mathbf{K}$ both for vectors and the positions of the edges of the Brillouin zone ($\mathrm{K}^\pm$), for clarity.
In the vicinity of $\pm\mathbf{K}$ points the Bloch states are predominantly made of the $d$-orbitals of metal atoms. The corresponding valence and conduction bands states of the bottom layer can be presented as \begin{align}
 \Psi^b_{\pm\mathbf{K},v}(\mathbf{r})=\frac{1}{\sqrt{N}}\sum_{\mathbf{R}_{mn}^{M,b}}
e^{\pm i\mathbf{K}\mathbf{R}_{mn}^{M,b}}Y_{2,\pm 2}(\mathbf{r}-\mathbf{R}_{mn}^{M,b}),\\
 \Psi^b_{\pm\mathbf{K},c}(\mathbf{r})=\frac{1}{\sqrt{N}}\sum_{\mathbf{R}_{mn}^{M,b}}
 e^{\pm i\mathbf{K}\mathbf{R}_{mn}^{M,b}}Y_{2,0}(\mathbf{r}-\mathbf{R}_{mn}^{M,b}).
 \end{align}
Here $N$ is the normalization factor and $Y_{lm}(\mathbf{r}-\mathbf{R})$ is the value of the $lm$-th atomic orbital placed at the point $\mathbf{R}$ and calculated at the point $\mathbf{r}$. The operator $\widehat{C}_3$, which generates $R_{2\pi/3}$ rotation of the vectors in space, transforms the corresponding Bloch functions as
\begin{widetext}
	\begin{align}
 \widehat{C}_3\Psi^b_{\pm\mathbf{K},v}(\mathbf{r})=&\frac{1}{\sqrt{N}}\sum_{\mathbf{R}_{mn}}e^{\pm i\mathbf{K}
 (\mathbf{R}_{mn}+\mathbf{t}_M)}Y_{2,\pm 2}(R^{-1}_{2\pi/3}\mathbf{r}-\mathbf{R}_{mn}-\mathbf{t}_M)=\nonumber \\
 =&e^{\mp4\pi i/3}\frac{1}{\sqrt{N}}\sum_{\mathbf{R}_{mn}}e^{\pm i\mathbf{K}
 (\mathbf{R}_{mn}+\mathbf{t}_M)}Y_{2,\pm 2}(\mathbf{r}-R_{2\pi/3}[\mathbf{R}_{mn}+\mathbf{t}_M])=\nonumber \\
 =&e^{\mp4\pi i/3}\frac{1}{\sqrt{N}}\sum_{\mathbf{R}_{m'n'}}e^{\pm iR_{2\pi/3}\mathbf{K}
 (\mathbf{R}_{m'n'}+\mathbf{t}_M)}Y_{2,\pm 2}(\mathbf{r}-\mathbf{R}_{m'n'}-\mathbf{t}_M)
=\nonumber \\
 =&e^{\mp4\pi i/3}e^{\mp i\mathbf{b}_2\mathbf{t}_M}\frac{1}{\sqrt{N}}\sum_{\mathbf{R}_{mn}}e^{\pm i\mathbf{K}
 (\mathbf{R}_{mn}+\mathbf{t}_M)}Y_{2,\pm 2}(\mathbf{r}-\mathbf{R}_{mn}-\mathbf{t}_M)=
\Psi^b_{\pm\mathbf{K},v}(\mathbf{r}), \\
\widehat{C}_3\Psi^b_{\pm\mathbf{K},c}(\mathbf{r})=&e^{\mp 2\pi i/3}\Psi^b_{\pm\mathbf{K},c}(\mathbf{r}).
 \end{align}
\end{widetext}
The phases, which appear under the transformation of the basis states of the bottom layer correspond to the notation in Ref.~\citenum{Liu2015}.

In addition, the crystal has inversion symmetry $I:\mathbf{r}\leftrightarrow -\mathbf{r}+2\mathbf{R}_I$, with the center of inversion in the point $\mathbf{R}_I=l\mathbf{e}_z/2$. This symmetry together with the time-reversal symmetry induces the restriction on the band structure of the crystal. In accordance to the Kramers theorem, all the bands of the crystal become doubly degenerated by spin. Therefore, it is convenient to define the second pair of basis states, associated with the top layer, using the above-mentioned symmetry operations $\Psi^t_{\pm\mathbf{K},n}(\mathbf{r})=\widehat{K}_0\widehat{I}\Psi^b_{\pm\mathbf{K},n}(\mathbf{r})$. Here $\widehat{K}_0$ and $\widehat{I}$ are complex conjugation and inversion symmetry operators, respectively.
Using the tight-binding representation of the basis states of the bottom layer we get
	\begin{widetext}
	\begin{align}
 \Psi^t_{\pm\mathbf{K},v}(\mathbf{r})=\widehat{K}_0\widehat{I}\Psi^b_{\pm\mathbf{K},v}(\mathbf{r})=
  &\frac{1}{\sqrt{N}}\sum_{\mathbf{R}_{mn}}e^{\mp i\mathbf{K}
 (\mathbf{R}_{mn}+\mathbf{t}_M)}Y^*_{2,\pm 2}(I^{-1}\mathbf{r}-\mathbf{R}_{mn}-\mathbf{t}_M)=\nonumber \\
 =&\frac{1}{\sqrt{N}}\sum_{\mathbf{R}_{mn}}e^{\mp i\mathbf{K}
 (\mathbf{R}_{mn}+\mathbf{t}_M)}Y_{2,\mp 2}(\mathbf{r}-I[\mathbf{R}_{mn}+\mathbf{t}_M])=\nonumber \\
 =&\frac{1}{\sqrt{N}}\sum_{\mathbf{R}_{m'n'}}e^{\mp iI\mathbf{K}
 (\mathbf{R}_{m'n'}+\mathbf{t}_X)}Y_{2,\mp 2}(\mathbf{r}-\mathbf{R}_{m'n'}-\mathbf{t}_X-l\mathbf{e}_z)
=\nonumber \\
 =&\frac{1}{\sqrt{N}}\sum_{\mathbf{R}_{mn}}e^{\pm i\mathbf{K}
 (\mathbf{R}_{mn}+\mathbf{t}_X)}Y_{2,\mp 2}(\mathbf{r}-\mathbf{R}_{mn}-\mathbf{t}_X-l\mathbf{e}_z), \\
\Psi^t_{\pm\mathbf{K},c}(\mathbf{r})=\widehat{K}_0\widehat{I}\Psi^b_{\pm\mathbf{K},c}(\mathbf{r})=&
\frac{1}{\sqrt{N}}\sum_{\mathbf{R}_{mn}}e^{\pm i\mathbf{K}
 (\mathbf{R}_{mn}+\mathbf{t}_X)}Y_{2,0}(\mathbf{r}-\mathbf{R}_{mn}-\mathbf{t}_X-l\mathbf{e}_z).
 \end{align}
\end{widetext}
The states satisfy the following transformation rules under rotation $\widehat{C}_3\Psi^t_{\pm\mathbf{K},v}(\mathbf{r})=\Psi^t_{\pm\mathbf{K},v}(\mathbf{r})$,
$\widehat{C}_3\Psi^t_{\pm\mathbf{K},c}(\mathbf{r})=e^{\pm 2\pi i/3}\Psi^t_{\pm\mathbf{K},c}(\mathbf{r})$,
with phases which are opposite to the phases of the basis states of the bottom layer
$\widehat{C}_3\Psi^b_{\pm\mathbf{K},v}(\mathbf{r})=
\Psi^b_{\pm\mathbf{K},v}(\mathbf{r})$,
$\Psi^b_{\pm\mathbf{K},c}(\mathbf{r})=e^{\mp 2\pi i/3}\Psi^b_{\pm\mathbf{K},c}(\mathbf{r})$. It leads to the fact that bilayer crystal can absorb the light with both circular polarizations in $\mathbf{K}$ point as well as in $-\mathbf{K}$ one. This feature is a consequence of inversion symmetry of the crystal. Finally, the mirror symmetry operator $\widehat{P}$ acts on the basis states as $\widehat{P}\Psi^\alpha_{\pm\mathbf{K},n}(\mathbf{r})=[\Psi^\alpha_{\pm\mathbf{K},n}(\mathbf{r})]^*=
\Psi^\alpha_{\mp\mathbf{K},n}(\mathbf{r})$, where $n=c,v$ and $\alpha=b,t$.

In further we focus on the states at the $\mathbf{K}$ point for brevity. We take into account the spin degree of freedom $s=\uparrow,\downarrow$ of electron excitations and introduce the following set of 8 basis states
\begin{align}
|\Psi_c^b,s\rangle=\Psi^b_{\mathbf{K},c}(\mathbf{r})|s\rangle, \quad
|\Psi_v^b,s\rangle=\Psi^b_{\mathbf{K},v}(\mathbf{r})|s\rangle, \\
|\Psi_c^t,s\rangle=\Psi^t_{\mathbf{K},c}(\mathbf{r})|s\rangle, \quad
|\Psi_v^t,s\rangle=\Psi^t_{\mathbf{K},v}(\mathbf{r})|s\rangle.
\end{align}
According to the $\mathbf{kp}$ method, developed for S-TMD multilayers \cite{MolasWS2,Arora2018,Slobodeniuk2019} the quasiparticles with the momentum $\mathbf{k}=k_x\mathbf{e}_x+k_y\mathbf{e}_y$ at the $\mathbf{K}$ point are described by the matrix elements $\langle\Psi_n^\alpha,s|\widehat{H}|\Psi_{n'}^{\alpha'},s'\rangle$ of the one-particle Hamiltonian
\begin{widetext}
\begin{equation}
H(\mathbf{r}) = \frac{\mathbf{\widehat{p}}^2}{2m_0}+ U^b(\mathbf{r}) +
U^t(\mathbf{r})+
\frac{\hbar}{4m_0^2c^2}\big[\nabla U^b(\mathbf{r}),\mathbf{\widehat{p}}\big]\boldsymbol{\sigma}+
\frac{\hbar}{4m_0^2c^2}\big[\nabla U^t(\mathbf{r}),\mathbf{\widehat{p}}\big]\boldsymbol{\sigma}+
\frac{\hbar}{m_0}\mathbf{k\widehat{p}}.
\end{equation}
\end{widetext}
Here $m_0$ is electron's mass, $c$ -- speed of light, $\hbar$ -- Planck's constant, $\boldsymbol{\sigma}=(\sigma_x, \sigma_y, \sigma_z)$ are Pauli matrices and $\mathbf{\widehat{p}}=-i\hbar\nabla$ is the momentum operator. The first term of the Hamiltonian defines the kinetic energy of an electron which propagates in the crystal field of the bottom $U^b(\mathbf{r})$ and top $U^t(\mathbf{r})$ layers of the bilayer. The next two terms describe the spin-orbital interaction in the system, induced by the potentials $U^b(\mathbf{r})$ and $U^t(\mathbf{r})$, respectively. The last $\mathbf{kp}$ term couples valence and conduction bands. This coupling is supposed to be small and we omit its effects for the current study. The detailed analysis of the impact of the  $\mathbf{kp}$ term can be found in Refs.~\citenum{Arora2018,Slobodeniuk2019}.

We consider first the matrix elements of the states of bottom layer. We present the Hamiltonian as
\begin{equation}
H(\mathbf{r})=H_0^b(\mathbf{r}) + H^t_{int}(\mathbf{r}),
\end{equation}
where we introduced the Hamiltonian of the bottom monolayer
\begin{equation}
H_0^b(\mathbf{r})=\frac{\mathbf{\widehat{p}}^2}{2m_0}+ U^b(\mathbf{r}) +
\frac{\hbar}{4m_0^2c^2}\big[\nabla U^b(\mathbf{r}),\mathbf{\widehat{p}}\big]\boldsymbol{\sigma},
\end{equation}
and the term which affects the motion of quasiparticles of the bottom layer by the crystal field of the top layer
\begin{equation}
H^t_{int}(\mathbf{r})=U^t(\mathbf{r})+\frac{\hbar}{4m_0^2c^2}\big[\nabla U^t(\mathbf{r}),\mathbf{\widehat{p}}\big]\boldsymbol{\sigma}.
\end{equation}
The Hamiltonian $H_0^b(\mathbf{r})$ has the diagonal matrix elements,which are nothing but the position of the
conduction and valence bands in monolayer
\begin{align}
\langle\Psi_v^b,s|H_0^b(\mathbf{r})|\Psi_v^b,s\rangle=E_v+\sigma_s\Delta_v/2, \\
\langle\Psi_c^b,s|H_0^b(\mathbf{r})|\Psi_c^b,s\rangle=E_c+\sigma_s\Delta_c/2,
\end{align}
Here $E_v$ and $E_c$ are positions of the valence and conduction bands without spin splitting, $\Delta_v$ and $\Delta_c$ are their spin splittings, and $\sigma_s=+1(-1)$ for $s=\uparrow(\downarrow)$ states respectively.
Note that in $\mathbf{K}$ point $\Delta_v$ is always positive, while $\Delta_c$ can be negative (bright type of S-TMD) and positive (darkish type of S-TMD). 

The dominant contribution from of $H^t_{int}(\mathbf{r})$ is the diagonal matrix elements in spin space
\begin{align}
\langle\Psi_v^b,s|H^t_{int}(\mathbf{r})|\Psi_v^b,s\rangle=\delta E_v+\sigma_s\delta\Delta_v/2, \\
\langle\Psi_c^b,s|H^t_{int}(\mathbf{r})|\Psi_c^b,s\rangle=\delta E_c+\sigma_s\delta\Delta_c/2.
\end{align}
We suppose that the corrections to the splitting are small $|\Delta_c|\gg|\delta \Delta_c|$, $|\Delta_v|\gg|\delta \Delta_v|$, and the type of S-TMD bilayer remains the same as the type of its constituents. $H^t_{int}(\mathbf{r})$ term has also non-zero matrix element between valence and conduction bands of the same layer and opposite spins (see Refs.~\citenum{Ochoa2013,Slobodeniuk2016} for details). However, these matrix elements give the negligibly small contribution to the energies of the bands, proportional to $\sim 1/(E_c-E_v)$. Therefore, we omit them from the study.

The matrix elements between the states of the top layer can be calculated in the same way. We present the total Hamiltonian as
\begin{equation}
H(\mathbf{r})=H_0^t(\mathbf{r}) + H^b_{int}(\mathbf{r}),
\end{equation}
where the first term 
\begin{equation}
H_0^t(\mathbf{r})=\frac{\mathbf{\widehat{p}}^2}{2m_0}+ U^t(\mathbf{r}) +
\frac{\hbar}{4m_0^2c^2}\big[\nabla U^t(\mathbf{r}),\mathbf{\widehat{p}}\big]\boldsymbol{\sigma},
\end{equation}
is the Hamiltonian of the top monolayer, while the second term 
\begin{equation}
H^b_{int}(\mathbf{r})=U^b(\mathbf{r})+
\frac{\hbar}{4m_0^2c^2}\big[\nabla U^b(\mathbf{r}),\mathbf{\widehat{p}}\big]\boldsymbol{\sigma}, 
\end{equation}
affects the motion of quasiparticles of the top layer by the crystal field of the bottom one. With the help of Kramers theorem and the latter result one can immediately get the answer for the matrix elements of the above-mentioned terms
\begin{align}
\langle\Psi_v^t,s|H_0^t(\mathbf{r})|\Psi_v^t,s\rangle&=E_v-\sigma_s\Delta_v/2, \\
\langle\Psi_c^t,s|H_0^t(\mathbf{r})|\Psi_c^t,s\rangle&=E_c-\sigma_s\Delta_c/2, \\
\langle\Psi_v^t,s|H^b_{int}(\mathbf{r})|\Psi_v^t,s\rangle&=\delta E_v-\sigma_s\delta\Delta_v/2, \\
\langle\Psi_c^t,s|H^b_{int}(\mathbf{r})|\Psi_c^t,s\rangle&=\delta E_c-\sigma_s\delta\Delta_c/2.
\end{align}
Note that the sign before spin-splitting terms for the states of the top layer is opposite to the sign of the same terms of the bottom layer. This is the manifestation of the double degeneracy by spin of all the bands of the bilayer.

Finally, we calculate the matrix elements of the Hamiltonian between the states of the different layers. Namely, we evaluate the following interlayer matrix elements  
$\langle\Psi_n^t,s|H(\mathbf{r})|\Psi_{n'}^b,s'\rangle$,
where $n,n'=c,v$. The other matrix elements can be obtained by complex conjugation of the considered ones. We present the Hamiltonian in the following way
\begin{equation}
H(\mathbf{r})=
H^t_0(\mathbf{r})+H^b_0(\mathbf{r})-\frac{\mathbf{\widehat{p}}^2}{2m_0}.
\end{equation}
and suppose the orthogonality of the states from the opposite layers
$\langle\Psi_n^t,s|\Psi_{n'}^b,s'\rangle=0$. Then 
\begin{equation}
\langle\Psi_n^t,s|H(\mathbf{r})|\Psi_{n'}^b,s'\rangle=
-\frac{1}{2m_0}\langle\Psi_n^t,s|\mathbf{\widehat{p}}^2|\Psi_{n'}^b,s'\rangle.
\end{equation}

\begin{figure}[!t]
    \centering
    \includegraphics[width=0.8\linewidth]{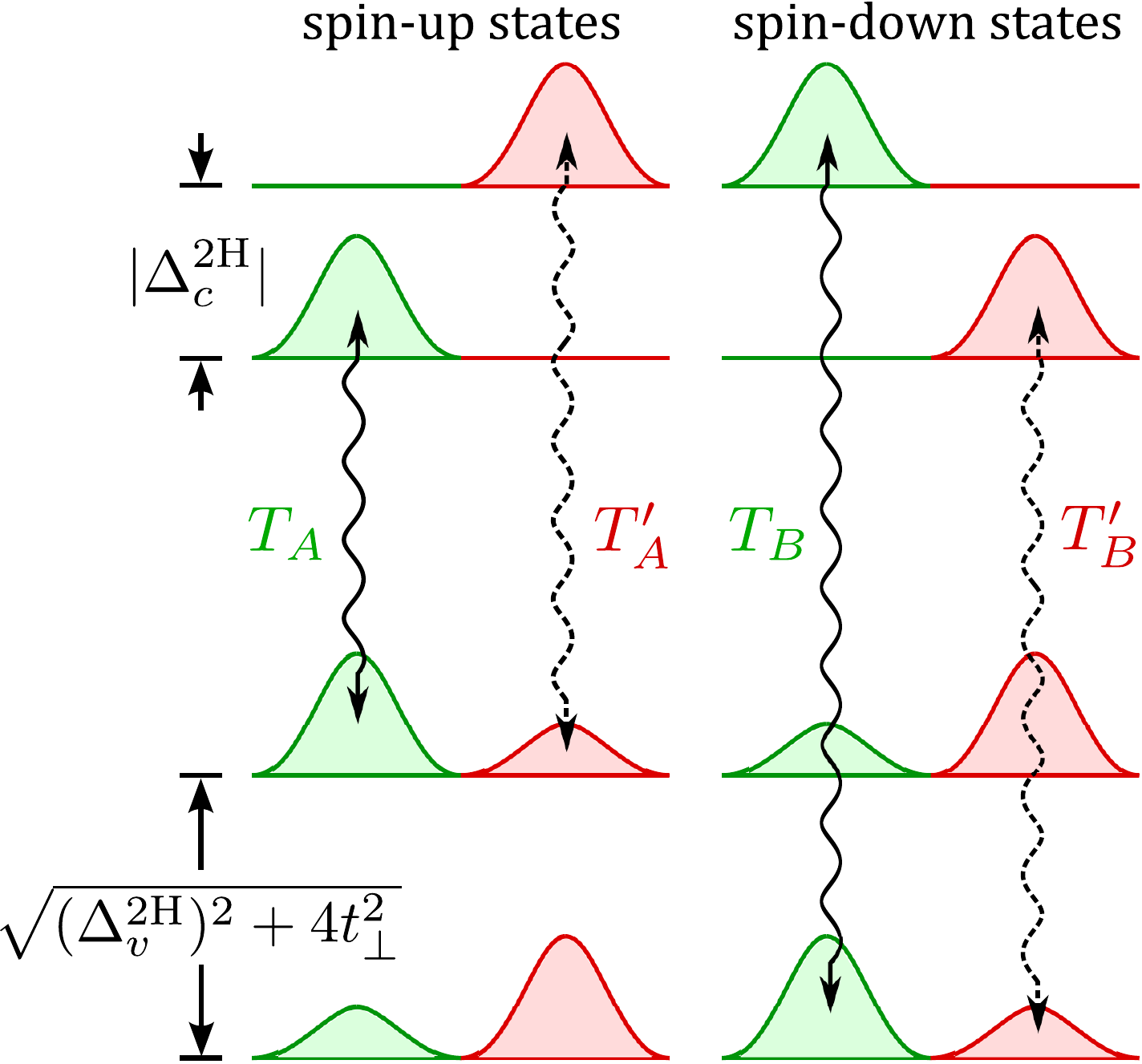}
    \caption{ The bands positions and optical transitions in the $+\mathbf{K}$ point of 2H stacked MoS$_2$ bilayer. Left/right side represents the spin-up/spin-down states in the system. Green and red convexes represent the conduction and valence band states associated with optical transitions active in the $\sigma^+$  and $\sigma^-$  polarizations, respectively. The green/red color also denotes the bottom/top layer. Solid (dashed) wavy arrows indicates optical transitions $T_A$, $T_B$ ($T_A'$ , $T_B'$ ) due to the intralayer  A and B (interlayer A$'$ and B$'$) excitons. $|\Delta^\textrm{2H}_c|$  and $\sqrt{(\Delta^\textrm{2H}_v)^2+4t_\perp^2}$  denote the splitting in the conduction ($c$) and the valence ($v$) bands, respectively.}
    \label{fig:Bilayer_2H}
\end{figure}

The $\mathbf{\widehat{p}}^2$ operator is a spin singlet, hence the matrix elements are diagonal in spin space $\langle\Psi_n^t,s|\mathbf{\widehat{p}}^2|\Psi_{n'}^b,s'\rangle=
\delta_{ss'}\langle\Psi_n^t|\mathbf{\widehat{p}}^2|\Psi_{n'}^b\rangle$, where
$|\Psi_n^\alpha\rangle=\Psi_{\mathbf{K},n}^\alpha(\mathbf{r})$.
Using transformation properties of the basis states under $C_3$ rotation 
\begin{widetext}
\begin{align}
\langle\Psi_v^t|\mathbf{\widehat{p}}^2|\Psi_c^b\rangle&=
\langle\Psi_v^t|\widehat{C}^{-1}_3\widehat{C}_3\mathbf{\widehat{p}}^2\widehat{C}^{-1}_3\widehat{C}_3|\Psi_c^b\rangle=
e^{-2\pi i/3}\langle\Psi_v^t|\mathbf{\widehat{p}}^2|\Psi_c^b\rangle=0, \\
\langle\Psi_c^t|\mathbf{\widehat{p}}^2|\Psi_c^b\rangle&=
\langle\Psi_c^t|\widehat{C}^{-1}_3\widehat{C}_3\mathbf{\widehat{p}}^2\widehat{C}^{-1}_3\widehat{C}_3|\Psi_c^b\rangle=
e^{-4\pi i/3}\langle\Psi_c^t|\mathbf{\widehat{p}}^2|\Psi_c^b\rangle=0, \\
\langle\Psi_v^t|\mathbf{\widehat{p}}^2|\Psi_v^b\rangle&=
\langle\Psi_v^t|\widehat{C}^{-1}_3\widehat{C}_3\mathbf{\widehat{p}}^2\widehat{C}^{-1}_3\widehat{C}_3|\Psi_v^b\rangle=
\langle\Psi_v^t|\mathbf{\widehat{p}}^2|\Psi_v^b\rangle=t_\perp\neq0, \\
\langle\Psi_c^t|\mathbf{\widehat{p}}^2|\Psi_v^b\rangle&=
\langle\Psi_c^t|\widehat{C}^{-1}_3\widehat{C}_3\mathbf{\widehat{p}}^2\widehat{C}^{-1}_3\widehat{C}_3|\Psi_v^b\rangle=
e^{-2\pi i/3}\langle\Psi_c^t|\mathbf{\widehat{p}}^2|\Psi_v^b\rangle=0,
\end{align}
\end{widetext}
we get that only one matrix element is non-zero. The $P$ symmetry of the crystal dictates that the parameter $t_\perp$ is a real number $\textrm{Im}[t_\perp]=0$. This parameter couples the valence bands of the same spin of the top and bottom layers, mixes them and forms the new type of valence band states, presented in Fig.~\ref{fig:Bilayer_2H} as two-convex structure. It indicates that the valence band Bloch state in the  $+\mathbf{K}$ point of the bilayer is a superposition of the corresponding valence band Bloch states of the bottom and top layers of the same spin. On the contrary, the single-convex representation of the conduction band Bloch states in the $+\mathbf{K}$ point of the bilayer indicates that the corresponding conduction band states of the top and bottom layers are not mixed. As a result, the dipole matrix elements between the new valence band states and conduction band states of the top as well as of the bottom layers become non-zero. It causes to doubling of number of possible optical transition in the system.

Summarizing the aforementioned calculations we conclude that there are four optical transitions in 2H-stacked bilayer:
$A$ and $B$ intense optical transitions (which form intralayer exciton complexes) and weak optical transitions 
$A'$ and $B'$(which form inter-layer exciton complexes). 
The energies of these excitons are 
\begin{widetext}
\begin{align}
E_A=
-\mathcal{E}_{A}+E^\textrm{2H}_c-E^\textrm{2H}_v+\frac{\Delta^\textrm{2H}_c}{2}-\frac{\sqrt{(\Delta^\textrm{2H}_v)^2+4t_\perp^2}}{2}, \\
E_B=-\mathcal{E}_{B}+E^\textrm{2H}_c-E^\textrm{2H}_v-
\frac{\Delta^\textrm{2H}_c}{2}+\frac{\sqrt{(\Delta_v^\textrm{2H})^2+4t_\perp^2}}{2}, \\
E_{A'}=-\mathcal{E}_{A'}+E^\textrm{2H}_c-E^\textrm{2H}_v-
\frac{\Delta^\textrm{2H}_c}{2}-\frac{\sqrt{(\Delta_v^\textrm{2H})^2+4t_\perp^2}}{2}, \\
E_{B'}=-\mathcal{E}_{B'}+E^\textrm{2H}_c-E^\textrm{2H}_v+
\frac{\Delta^\textrm{2H}_c}{2}+\frac{\sqrt{(\Delta^\textrm{2H}_v)^2+4t_\perp^2}}{2}.
\end{align}
\end{widetext}
Here we introduced he absolute values of the binding energies of corresponding excitons $\mathcal{E}_{A},\mathcal{E}_{B},\mathcal{E}_{A'}, \mathcal{E}_{B'}$, and we 
the short notations $E^\textrm{2H}_c=E_c+\delta E_c$, $E^\textrm{2H}_v=E_v+\delta E_v$, $\Delta_c^\textrm{2H}=\Delta_c+\delta \Delta_c$ and $\Delta_v^\textrm{2H}=\Delta_v+\delta \Delta_v$.
The sketch of the bands position in MoS$_2$ bilayer with 2H-stacking is presented in Fig.~\ref{fig:Bilayer_2H}.

For the particular case of MoS$_2$ the splitting in conduction band is supposed to be much smaller than the splitting in valence band $|\Delta^\textrm{2H}_c|\ll|\Delta^\textrm{2H}_v|$, and the binding energies of $A$ and $B$ excitons are considered to be equal $\mathcal{E}_A=\mathcal{E}_B$. In this approximation we have the following result $\Delta^\textrm{2H}_{A-B}=E_B-E_A\approx \sqrt{(\Delta_v^\textrm{2H})^2+4t_\perp^2}$.  

Note that the intralayer and interlayer optical transitions in the same $\mathbf{K}$ (or $-\mathbf{K}$) point are characterized by opposite circular polarizations and $g$-factors of corresponding excitons (see Refs.~\citenum{Slobodeniuk2019,Arora2018} for details).

\section{Bilayer with $0^\circ$-angle alignment}
\label{app:0_stacking}

In order to compare the results of the measurements presented in the main text we describe the optical properties of the bilayer S-TMD with zero-angle alignment (in further ``bilayer'') in the similar way as it was done for 2H-stacked bilayer. Again, we consider the bilayer as a pile of two monolayers (top and bottom), placed in parallel to $xy$ plane. We assume the positions of metal and chalcogen atoms of the bottom layer are the same as in the previous section
\begin{align}
\mathbf{R}_{mn}^{M,b}&=\mathbf{R}_{mn}+\mathbf{t}_M, \\
\mathbf{R}_{mn}^{X_\pm,b}&=\mathbf{R}_{mn}+\mathbf{t}_X\pm \eta\mathbf{e}_z,
\end{align}
The top lattice of the bilayer can be obtained from the bottom one as a  result of two consequent shifts: along $z$
direction on distance $l$ (which is not equal to the distance $l$ for the 2H-stacked bilayer) and then along $y$ direction on distance $a_0/\sqrt{3}$. Then, the position of the metal and chalcogen atoms of the top lattice can be presented as
\begin{align}
\mathbf{R}_{mn}^{M,t}&=\mathbf{R}_{mn}+l\mathbf{e}_z+\mathbf{a}_1, \\
\mathbf{R}_{mn}^{X_\pm,t}&=\mathbf{R}_{mn}+l\mathbf{e}_z+\mathbf{t}_M\pm \eta\mathbf{e}_z.
\end{align}
Note that the half of the chalcogen and half of metal atoms in this bilayer are aligned in $z$-direction. This type of stacking for hexagonal lattices is called Bernal or AB-stacking. The unit cell of the considering bilayer contains twice more atoms than in monolayer. The positions of metal and chalcogen atoms within the unit cell are defined by vectors $\{\mathbf{t}_M, \mathbf{a}_1+l\mathbf{e}_z\}$ and $\{\mathbf{t}_X\pm \eta\mathbf{e}_z, \mathbf{t}_M\pm\eta\mathbf{e}_z +l\mathbf{e}_z\}$, respectively.

Note that the considering lattice has neither in-plane mirror symmetry (like AA-stacked case) nor inversion symmetry (like 2H-stacked bilayer). It possesses only $C_3$ rotation symmetry (with $Oz$ line as a rotational axis) and mirror symmetry $P:x\leftrightarrow -x$ (the mirror's plane is $yz$-plane). Again the crystal has the same hexagonal Brillouin zone as the Brillouin zone of the bottom layer. Hence, we choose the same basis states for valence and conduction bands in the $\pm\mathbf{K}$ points of the bottom layer as we have in the previous section
\begin{align}
 \Psi^b_{\pm\mathbf{K},v}(\mathbf{r})=\frac{1}{\sqrt{N}}\sum_{\mathbf{R}_{mn}^{M,b}}
e^{\pm i\mathbf{K}\mathbf{R}_{mn}^{M,b}}Y_{2,\pm 2}(\mathbf{r}-\mathbf{R}_{mn}^{M,b}),\\
 \Psi^b_{\pm\mathbf{K},c}(\mathbf{r})=\frac{1}{\sqrt{N}}\sum_{\mathbf{R}_{mn}^{M,b}}
 e^{\pm i\mathbf{K}\mathbf{R}_{mn}^{M,b}}Y_{2,0}(\mathbf{r}-\mathbf{R}_{mn}^{M,b}).
 \end{align}
We define the Bloch states of the top lattice in the same way
\begin{align}
 \Psi^t_{\pm\mathbf{K},v}(\mathbf{r})=\frac{1}{\sqrt{N}}\sum_{\mathbf{R}_{mn}^{M,t}}
e^{\pm i\mathbf{K}\mathbf{R}_{mn}^{M,t}}Y_{2,\pm 2}(\mathbf{r}-\mathbf{R}_{mn}^{M,t}),\\
 \Psi^t_{\pm\mathbf{K},c}(\mathbf{r})=\frac{1}{\sqrt{N}}\sum_{\mathbf{R}_{mn}^{M,t}}
 e^{\pm i\mathbf{K}\mathbf{R}_{mn}^{M,t}}Y_{2,0}(\mathbf{r}-\mathbf{R}_{mn}^{M,t}).
 \end{align}
They have the corresponding transformation rules $\widehat{C}_3\Psi^t_{\pm\mathbf{K},v}(\mathbf{r})=
e^{\pm 2\pi i/3}\Psi^t_{\pm\mathbf{K},v}(\mathbf{r})$ and
$\widehat{C}_3\Psi^t_{\pm\mathbf{K},c}(\mathbf{r})=\Psi^t_{\pm\mathbf{K},c}(\mathbf{r})$.
After rotation the basis states of the top layer get the phases which deviates from the corresponding phases of the basis states of the bottom layer. This difference is the result of the shift in $y$ direction of the atoms of the top lattice with respect to the bottom one. Despite this difference, the optical transition rules are not changed for each separate layer of bilayer. Namely the top and bottom layers in $\pm\mathbf{K}$ points absorb the $\sigma^\pm$ circular polarized light, respectively. These optical properties can be understood as a consequence of time-reversal symmetry which couples $\mathbf{K}$ and $-\mathbf{K}$ points of bilayer. This feature demonstrates the significant difference in optical properties of $0^\circ$- and $60^\circ$-aligned S-TMD bilayers. The mirror symmetry transformation $\widehat{P}$ also couples $\pm\mathbf{K}$ points $\widehat{P}\Psi^\alpha_{\pm\mathbf{K},n}(\mathbf{r})=[\Psi^\alpha_{\pm\mathbf{K},n}(\mathbf{r})]^*=
\Psi^\alpha_{\mp\mathbf{K},n}(\mathbf{r})$, where $n=c,v$ and $\alpha=b,t$.

Like in previous section we focus on the states at the $\mathbf{K}$ point, take into account the spin degrees
of freedom and introduce the basis states 
\begin{align}
|\Psi_c^b,s\rangle=\Psi^b_{\mathbf{K},c}(\mathbf{r})|s\rangle, \quad
|\Psi_v^b,s\rangle=\Psi^b_{\mathbf{K},v}(\mathbf{r})|s\rangle, \\
|\Psi_c^t,s\rangle=\Psi^t_{\mathbf{K},c}(\mathbf{r})|s\rangle, \quad
|\Psi_v^t,s\rangle=\Psi^t_{\mathbf{K},v}(\mathbf{r})|s\rangle.
\end{align}
We use the $\mathbf{kp}$ approach and calculate the matrix elements $\langle\Psi_n^\alpha,s|H(\mathbf{r})|\Psi_{n'}^{\alpha'},s'\rangle$ of the one-particle Hamiltonian
\begin{widetext}
\begin{equation}
H(\mathbf{r}) = \frac{\mathbf{\widehat{p}}^2}{2m_0}+ U^b(\mathbf{r}) +
U^t(\mathbf{r})+
\frac{\hbar}{4m_0^2c^2}\big[\nabla U^b(\mathbf{r}),\mathbf{\widehat{p}}\big]\boldsymbol{\sigma}+
\frac{\hbar}{4m_0^2c^2}\big[\nabla U^t(\mathbf{r}),\mathbf{\widehat{p}}\big]\boldsymbol{\sigma}+
\frac{\hbar}{m_0}\mathbf{k\widehat{p}}.
\end{equation}
\end{widetext}
All the terms in this Hamiltonian has the same meaning as in previous section. We also omit
the $\mathbf{kp}$ term, since we are focused on the optical transitions exactly in $\pm\mathbf{K}$
points for clarity. 

Let us consider the states of bottom layer. We present the Hamiltonian in the following form
\begin{equation}
H(\mathbf{r})=H_0^b(\mathbf{r}) + H^t_{int}(\mathbf{r}),
\end{equation}
where we introduced the Hamiltonian of the bottom monolayer
\begin{equation}
H_0^b(\mathbf{r})=\frac{\mathbf{\widehat{p}}^2}{2m_0}+ U^b(\mathbf{r}) +
\frac{\hbar}{4m_0^2c^2}\big[\nabla U^b(\mathbf{r}),\mathbf{\widehat{p}}\big]\boldsymbol{\sigma},
\end{equation}
and the term which affects the motion of quasiparticles of the bottom layer by the crystal field of the top layer
\begin{equation}
H^t_{int}(\mathbf{r})=U^t(\mathbf{r})+\frac{\hbar}{4m_0^2c^2}\big[\nabla U^t(\mathbf{r}),\mathbf{\widehat{p}}\big]\boldsymbol{\sigma}.
\end{equation}
The Hamiltonian $H_0^b(\mathbf{r})$ has the diagonal matrix elements
\begin{align}
\langle\Psi_v^b,s|H_0^b(\mathbf{r})|\Psi_v^b,s\rangle=E_v+\sigma_s\Delta_v/2, \\
\langle\Psi_c^b,s|H_0^b(\mathbf{r})|\Psi_c^b,s\rangle=E_c+\sigma_s\Delta_c/2,
\end{align}
where all the parameters have the same meaning as in the previous section.
The dominating contribution from of $H^t_{int}(\mathbf{r})$ has also diagonal matrix elements
\begin{align}
\langle\Psi_v^b,s|H^t_{int}(\mathbf{r})|\Psi_v^b,s\rangle=\delta E^t_v+\sigma_s\delta\Delta^t_v/2, \\
\langle\Psi_c^b,s|H^t_{int}(\mathbf{r})|\Psi_c^b,s\rangle=\delta E^t_c+\sigma_s\delta\Delta^t_c/2,
\end{align}
This term also has non-zero matrix element between valence and conduction bands of opposite spins. As in previous case they give a negligibly small contribution to the energy of the bands, proportional to $\sim 1/(E_c-E_v)$. Therefore, we omit these matrix element from the current study.

\begin{figure}[!t]
    \centering
    \includegraphics[width=\linewidth]{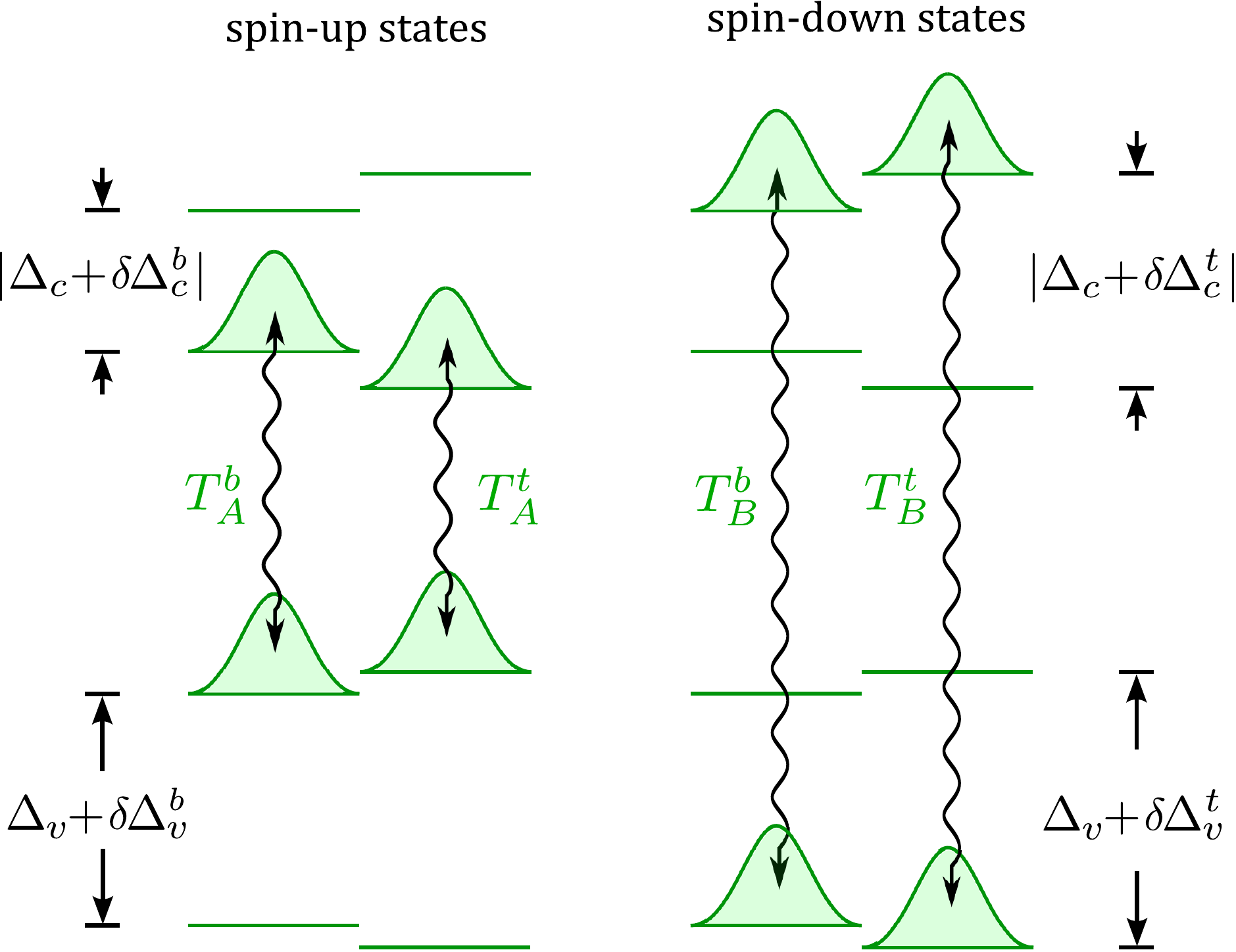}
    \caption{ The bands positions and optical transitions in the $+\mathbf{K}$ point of MoS$_2$ bilayer with $0^\circ$-angle alignment.
    Left/right side represents the spin-up/spin-down states in the system. Green convexes represent the conduction and valence band states associated with optical transitions active in the $\sigma^+$ polarization. Wavy arrows indicates optical transitions $T_A^b$, $T_A^t$, $T_B^b$, $T_B^t$  due to the intralayer A and B excitons in the botom ($b$) and top ($t$) layers. $|\Delta_v+\delta \Delta_v^b|$ ($|\Delta_v+\delta \Delta_v^t|$) and $|\Delta_c+\delta \Delta_c^b|$ ($|\Delta_c+\delta \Delta_c^t|$) denote the splitting in the conduction and valence bands of the bottom(top) layer, respectively.}
    \label{fig:Bilayer_0}
\end{figure}

The matrix elements between the states of the top layer can be calculated in the same way. Namely we present the Hamiltonian as
\begin{equation}
H(\mathbf{r})=H_0^t(\mathbf{r}) + H^b_{int}(\mathbf{r}),
\end{equation}
with the Hamiltonian of the top monolayer
\begin{equation}
H_0^t(\mathbf{r})=\frac{\mathbf{\widehat{p}}^2}{2m_0}+ U^t(\mathbf{r}) +
\frac{\hbar}{4m_0^2c^2}\big[\nabla U^t(\mathbf{r}),\mathbf{\widehat{p}}\big]\boldsymbol{\sigma},
\end{equation}
and the term which affects the motion of quasiparticles of the top layer by the crystal field of the bottom one
\begin{equation}
H^b_{int}(\mathbf{r})=U^b(\mathbf{r})+
\frac{\hbar}{4m_0^2c^2}\big[\nabla U^b(\mathbf{r}),\mathbf{\widehat{p}}\big]\boldsymbol{\sigma}.
\end{equation}
Again, only diagonal matrix elements of the $H_0^t(\mathbf{r})$ are nonzero
\begin{align}
\langle\Psi_v^t,s|H_0^t(\mathbf{r})|\Psi_v^t,s\rangle=E_v+\sigma_s\Delta_v/2, \\
\langle\Psi_c^t,s|H_0^t(\mathbf{r})|\Psi_c^t,s\rangle=E_c+\sigma_s\Delta_c/2,
\end{align}
The dominating contribution from of $H^b_{int}(\mathbf{r})$ has also diagonal matrix elements
\begin{align}
\langle\Psi_v^t,s|H^b_{int}(\mathbf{r})|\Psi_v^t,s\rangle=\delta E^b_v+\sigma_s\delta\Delta^b_v/2, \\
\langle\Psi_c^t,s|H^b_{int}(\mathbf{r})|\Psi_c^t,s\rangle=\delta E^b_c+\sigma_s\delta\Delta^b_c/2.
\end{align}
Note that due to the absence of the additional symmetry of the crystal (like inverse symmetry for 2H-stacked bilayers, or mirror symmetry for AA-stacked bilayer) we cannot find the relations between the parameters $\{\delta E^b_v,\delta\Delta^b_v,\delta E^b_c,\delta\Delta^b_c\}$ and $\{\delta E^t_v,\delta\Delta^t_v,\delta E^t_c,\delta\Delta^t_c\}$. Therefore, according to the symmetry analysis, we consider these parameters as independent ones.

Finally, we calculate the interlayer matrix elements $\langle\Psi_n^t,s|H(\mathbf{r})|\Psi_{n'}^b,s'\rangle$.
Supposing the orthogonality of the states from the opposite layers and repeating the idea of the calculation from
the previous section one gets
\begin{equation}
\langle\Psi_n^t,s|H(\mathbf{r})|\Psi_{n'}^b,s'\rangle=
-\frac{1}{2m_0}\langle\Psi_n^t,s|\mathbf{\widehat{p}}^2|\Psi_{n'}^b,s'\rangle.
\end{equation}

Since the $\mathbf{\widehat{p}}^2$ operator is a spin singlet, the matrix elements are diagonal in spin subspace $\langle\Psi_n^t,s|\mathbf{\widehat{p}}^2|\Psi_{n'}^b,s'\rangle=\delta_{ss'}\langle\Psi_n^t|\mathbf{\widehat{p}}^2|\Psi_{n'}^b\rangle$. Here we introduce the notation $|\Psi_n^\alpha\rangle=\Psi_{\mathbf{K},n}^\alpha(\mathbf{r})$.
Using transformation properties of the basis states under $C_3$ rotation we get
\begin{widetext}
\begin{align}
\langle\Psi_v^t|\mathbf{\widehat{p}}^2|\Psi_c^b\rangle&=
\langle\Psi_v^t|\widehat{C}^{-1}_3\widehat{C}_3\mathbf{\widehat{p}}^2\widehat{C}^{-1}_3\widehat{C}_3|\Psi_c^b\rangle=
e^{-4\pi i/3}\langle\Psi_v^t|\mathbf{\widehat{p}}^2|\Psi_c^b\rangle=0, \\
\langle\Psi_c^t|\mathbf{\widehat{p}}^2|\Psi_c^b\rangle&=
\langle\Psi_c^t|\widehat{C}^{-1}_3\widehat{C}_3\mathbf{\widehat{p}}^2\widehat{C}^{-1}_3\widehat{C}_3|\Psi_c^b\rangle=
e^{-2\pi i/3}\langle\Psi_c^t|\mathbf{\widehat{p}}^2|\Psi_c^b\rangle=0, \\
\langle\Psi_v^t|\mathbf{\widehat{p}}^2|\Psi_v^b\rangle&=
\langle\Psi_v^t|\widehat{C}^{-1}_3\widehat{C}_3\mathbf{\widehat{p}}^2\widehat{C}^{-1}_3\widehat{C}_3|\Psi_v^b\rangle=
e^{-2\pi i/3}\langle\Psi_v^t|\mathbf{\widehat{p}}^2|\Psi_v^b\rangle=0, \\
\langle\Psi_c^t|\mathbf{\widehat{p}}^2|\Psi_v^b\rangle&=
\langle\Psi_c^t|\widehat{C}^{-1}_3\widehat{C}_3\mathbf{\widehat{p}}^2\widehat{C}^{-1}_3\widehat{C}_3|\Psi_v^b\rangle=
\langle\Psi_c^t|\mathbf{\widehat{p}}^2|\Psi_v^b\rangle=\tau\neq0.
\end{align}
\end{widetext}
The $P$ symmetry of the crystal dictates that the parameter $\tau$ is a real number $\textrm{Im}[\tau]=0$. This parameter couples the valence band of the bottom layer with the conduction band of the top layer with the same spin state. It induces the negligibly small correction to the energies of the corresponding bands, proportional to $\sim 1/(E_c-E_v)$. Therefore, we omit such terms in our consideration.

Summarizing the aforementioned calculations we conclude that the states of the top and bottom layers are not mixed. Therefore, the considered bilayer has the same optical properties as a monolayer. Namely, this bilayer absorbs the $\sigma^\pm$ polarized light in the $\pm\mathbf{K}$ points that induces the intralayer dipole transitions $A^b, B^b$ and $A^t, B^t$ within the bottom and top layers, respectively. The energies of these excitons are 
\begin{widetext}
\begin{align}
E_{A^b}=-\mathcal{E}_{A^b}+E_c+\delta E^b_c-E_v-\delta E^b_v+
\frac{\Delta_c+\delta \Delta^b_c}{2}-\frac{\Delta_v+\delta \Delta^b_v}{2}, \\
E_{B^b}=-\mathcal{E}_{B^b}+E_c+\delta E^b_c-E_v-\delta E^b_v-
\frac{\Delta_c+\delta \Delta^b_c}{2}+\frac{\Delta_v+\delta \Delta^b_v}{2}, \\
E_{A^t}=-\mathcal{E}_{A^t}+E_c+\delta E^t_c-E_v-\delta E^t_v+
\frac{\Delta_c+\delta \Delta^t_c}{2}-\frac{\Delta_v+\delta \Delta^t_v}{2}, \\
E_{B^t}=-\mathcal{E}_{B^t}+E_c+\delta E^t_c-E_v-\delta E^t_v-
\frac{\Delta_c+\delta \Delta^t_c}{2}+\frac{\Delta_v+\delta \Delta^t_v}{2}.
\end{align}
\end{widetext}
Here $\mathcal{E}_{A^b},\mathcal{E}_{B^b},\mathcal{E}_{A^t}, \mathcal{E}_{B^t}$ are the absolute values of the binding energies of the corresponding excitons. The band positions for the MoS$_2$ bilayer with $0^\circ$-angle alignment is sketched in Fig.~\ref{fig:Bilayer_0}.

The energies of the top- and bottom-excitons of the same type ($A$ or $B$) deviate from each other in general case. The symmetry analysis cannot provide the value of the splitting between $A^b$ and $A^t$ ($B^b$ and $B^t$) exciton lines. However, as we do not observe any splittings of the $A$ and $B$ emission lines for 6$^\circ$ structure (see Fig.~\ref{fig:PL_temp}), we suppose that the $A^b$-$A^t$ and $B^b$-$B^t$ splittings are quite small and probably cannot be recognized in the experiment. As a result, we assume that the experimentally obtained $A$-$B$ energy distance equals $\Delta^{0^{\circ}}_{A-B}=(E_{B^b}+E_{B^t})/2-(E_{A^b}+E_{A^t})/2$. Similarly to the 2H-stacked MoS$_2$ bilayer, the splittings in conduction band are supposed to be much smaller than the splitting in valence band: $|\Delta_c+\delta \Delta^b_c|\ll\Delta_v+\delta \Delta^b_v$ and $|\Delta_c+\delta \Delta^t_c|\ll\Delta_v+\delta \Delta^t_v$, and the binding energies of $A$ and $B$ excitons are considered to be equal $\mathcal{E}_{A^b}=\mathcal{E}_{A^t}=\mathcal{E}_{B^b}=
\mathcal{E}_{B^t}$. It implies that 
$\Delta^{0^{\circ}}_{A-B}=(E_{B^b}+E_{B^t})/2-(E_{A^b}+E_{A^t})/2\approx\Delta_v+(\delta \Delta^b_v+\delta \Delta^t_v)/2$. Morevoer, if we assume that the $\delta \Delta^b_v\approx\delta \Delta^t_v\approx\delta \Delta_v$ (where $\delta \Delta_v$ is the corresponding parameter calculated for 2H-stacked BL), then $\Delta^{0^{\circ}}_{A-B}\approx\Delta^\textrm{2H}_v$.

\bibliographystyle{apsrev4-1}
\bibliography{biblio}

\end{document}